\documentclass[12pt]{article}
\usepackage{hyperref,color}
\usepackage{amsfonts}
\usepackage{latexsym}
\usepackage{amsmath,amssymb}
\usepackage{cite}

\newcommand{\bea}{\begin{eqnarray}}
\newcommand{\eea}{\end{eqnarray}}
\newcommand{\be}{\begin{equation}}
\newcommand{\ee}{\end{equation}}
\newcommand{\ba}{\begin{array}}
\newcommand{\ea}{\end{array}}

\def\nn{\nonumber}

\def\p{\partial}
\def\eps{\epsilon}

\numberwithin{equation}{section}

\def\sdelta{\slash\hspace{-6pt}\delta}

\def\cH{\mathcal{H}}
\def\cT{\mathcal{T}}
\def\cQ{\mathcal{Q}}
\def\cL{\mathcal{L}}
\def\cJ{\mathcal{J}}
\def\cG{\mathcal{G}}
\def\cW{\mathcal{W}}
\def\cM{\mathcal{M}}
\def\tr{\hbox{Tr}}
\def\sdelta{\slash\hspace{-6pt}\delta}

\definecolor{Wei}{rgb}{0.65,0.0,0}
\definecolor{Geo}{rgb}{0.1,0,0.75}

\begin{document}

\begin{center}\vspace{2cm}
{ \LARGE {{$\cW$ Symmetry and Integrability\vspace{6pt} \\
of Higher spin black holes}}}

\vspace{1cm}

Geoffrey Comp\`ere$^{*\diamond}$, and Wei Song$^*$

\vspace{0.8cm}

{\it  $*$Center for the Fundamental Laws of Nature, Harvard University,\\
Cambridge, MA, USA}

{\it  $\diamond$ Physique Th\'eorique et Math\'ematique, Universit\'e Libre de Bruxelles,\\
 Bruxelles, Belgium}

\vspace{0.5cm}
\today

\vspace{0.5cm}

\vspace{1.0cm}

\end{center}

\begin{abstract}

We obtain the asymptotic symmetry algebra of $sl(3,\mathbb R) \times sl(3,\mathbb R)$ Chern-Simons theory  with Dirichlet boundary conditions for fixed chemical potential. These boundary conditions are obeyed by higher spin black holes. For each embedding of $sl(2,\mathbb R)$ into $sl(3,\mathbb R)$, we show that the asymptotic symmetry group is independent of the chemical potential. On the one hand, starting from $AdS_3$ in the principal embedding, we show that the $\cW_3 \times \cW_3$ symmetry is preserved upon turning on perturbatively \mbox{spin 3} chemical potentials. On the other hand, starting from $AdS_3$ in the diagonal embedding, we show that the $\cW_3^{(2)} \times \cW_3^{(2)}$ symmetry is preserved upon turning on finite \mbox{spin 3/2} chemical potentials. We also make connections between the canonical Lagrangian formalism and integrability methods based on the $n=3$ KdV (Boussinesq) hierarchy.

\vspace{1cm}

\end{abstract}
\thispagestyle{empty}

\pagebreak
\setcounter{tocdepth}{2}

\tableofcontents

\section{Introduction}

In three spacetime dimensions, Vasiliev higher spin theory consists of an infinite tower of higher spins fields coupled to a scalar field, which is characterized by its mass labeled by a continuous parameter $\lambda$ \cite{Vasiliev:1995dn,Vasiliev:1996hn}.
The sector of pure higher spin theories can be formulated as a Chern-Simons gauge theory  \cite{Blencowe:1988gj,Bergshoeff:1989ns}. When there is no matter, the infinite tower of higher spin fields originating from the gauge algebra $hs(\lambda) \times hs(\lambda)$ can be truncated to a finite tower for any integer $N$ when the coupling is restricted as $\lambda = N$. The result is the $sl(N,\mathbb R) \times sl(N,\mathbb R)$ Chern-Simons theory.  In this paper, we study the simplest such higher spin theory for $N=3$ even though we expect that our analysis can be extended in a straightforward manner to more general cases.

One motivation for the present work is the conjectured holographic correspondence proposed by Gaberdiel and Gopakumar  \cite{Gaberdiel:2010pz}. The conjecture relates the Vasiliev theory to a large $N$ limit of $W_N$ minimal models at fixed 't Hooft coupling $\lambda$ where $\lambda$ is the deformation parameter of the $hs(\lambda)$ bulk algebra (see  \cite{Ahn:2011pv,Chang:2011mz,Gaberdiel:2011nt,Creutzig:2011fe,Gaberdiel:2012ku,Creutzig:2012ar,Chang:2013izp,Beccaria:2013wqa,Gaberdiel:2013vva} for further refinements and extensions of the original conjecture). Partition functions have been computed in \cite{Kraus:2011ds,Gaberdiel:2011zw}, and three point functions have been checked in \cite{Chang:2011mz,Ammon:2011ua,Papadodimas:2011pf,Chang:2011vka}. Other aspects of the duality have been investigated extensively \cite{Castro:2011zq,Gaberdiel:2011aa,Candu:2012jq,Candu:2012tr,Candu:2012ne,Candu:2013uya}, see \cite{Gaberdiel:2012uj} for a review.

In $sl(N,\mathbb R) \times sl(N,\mathbb R)$ Chern-Simons theory, the definition of a metric requires to define an embedding of $sl(2,\mathbb R)$ into the higher spin gauge algebra. An AdS$_3$ vacuum exists for each choice of embedding. Two particular embeddings can be defined for any $N$: the principal and diagonal embedding while more embeddings exist for $N > 3$.

The analysis of Brown-Henneaux \cite{Brown:1986nw} has been generalized to higher spin theories with asymptotic $AdS_3$ boundary conditions. Incidentally, the identification of asymptotic symmetries in Chern-Simons theory has first be obtained using the so-called Drinfeld-Sokolov Hamiltonian reduction \cite{Drinfeld:1981aa,Drinfeld:1985aa}. Following the Brown-Henneaux approach, the asymptotic symmetry algebra for asymptotically $AdS_3$ solutions in the principal embedding has been computed for the $hs(1/2)$ Chern-Simons gauge algebra, which resulted in the $\cW_\infty(1/2)$ non-linear algebra \cite{Henneaux:2010xg} (see also its supersymmetric extensions \cite{Henneaux:2012ny,Hanaki:2012yf}). It has been independently obtained for the $sl(N,\mathbb R)$ gauge algebra, which resulted in the $\cW_N$ algebra \cite{Campoleoni:2010zq}. Both results were generalized for the $hs(\lambda)$ algebra, which led to the $\cW_\infty(\lambda)$ asymptotic symmetry algebra \cite{Gaberdiel:2011wb,Campoleoni:2011hg} (see \cite{Feher:1992yx,Bouwknegt:1992wg, DickeyLectures} for a summary of results on $\cW$ algebras). The asymptotic symmetry group for asymptotically $AdS_3$ solutions in the diagonal embedding has also been computed for $sl(3,\mathbb R)$ in  \cite{Ammon:2011nk} with the Polyakov-Bershadsky $\cW^{(2)}_3$ algebra \cite{Polyakov:1989dm,Bershadsky:1990bg} as a result.
Note that here and in what follows, we will only discuss the classical version of $\cW$-algebras.

The fact that  the same solution leads to different asymptotic symmetry algebras might be confusing. In this paper, we first clarify that different embeddings are equivalent to imposing different choices of boundary conditions on an initial data slice.
We will indeed observe that in order to perform the canonical analysis, the initial data problem has to be defined as a part of the boundary conditions. For the principal embedding, the initial data amounts to the value of the two functions $\cL$ and $\cW$ that parameterize the spin 2 and spin 3 fields. For the diagonal embedding, more initial data is required. One can formulate this initial data either as the values of the spin 2, two spin 3/2 and spin 1 functions $\mathcal{T}, \mathcal{G}^{\pm},\cJ$ at the initial time, or as $\cL$, $\cW$ together with the first and second time derivative of $\cL$. The two initial data sets are related by field redefinitions. The choice of boundary conditions is also equivalent to a choice of quantization.  Indeed, it was already noticed that the Ward identities of either the $\cW_3$ or $\cW^{(2)}_3$ algebras appear as the zero curvature condition for the $sl(3,\mathbb R)$ Chern-Simons theory depending on the choice of quantization \cite{Bilal:1991cf,Bilal:1992ub}.

In three dimensions there are BTZ black holes \cite{Banados:1992wn}, which are locally a quotient of global $AdS_3$ \cite{Banados:1992gq}. Black holes carrying higher spin charges have also been found \cite{Gutperle:2011kf}. Their spacetime structure has been discussed \cite{Ammon:2011nk} and their thermodynamics has been investigated \cite{Gutperle:2011kf,Kraus:2012uf,Perez:2012cf,Campoleoni:2012hp,Perez:2013xi,Kraus:2013esi,deBoer:2013gz,Ferlaino:2013vga}. The phase structure of black holes was further explored in \cite{David:2012iu,Banerjee:2012aj,Chen:2012ba}. For more related work, see \cite{Ammon:2012wc} for a review. One important feature of such black holes is that consistency of thermodynamics requires that chemical potentials be functionals of higher spin charges. Boundary conditions admitting non-zero higher spin chemical potentials are therefore essential in order to include higher spin black holes as admissible solutions. In this paper, we will build up such boundary conditions and compute their asymptotic symmetry algebra.

It was argued \cite{Ammon:2011nk}  that black holes are RG flows between two distinct conformal field theories dual to $AdS_3$ vacua with distinct asymptotic symmetry algebras. In the case of the $sl(3,\mathbb R)$ gauge algebra, black holes with \mbox{spin $3$} chemical potentials $\mu,\bar \mu$ are indeed interpolating solutions between an $AdS_3$ of radius $l/2$ and an $AdS_3$ with radius $l$. It was then argued that since the asymptotic symmetry algebra in both $AdS_3$ geometries are different (respectively $\cW_3^{(2)} \times \cW_3^{(2)}$ and $\cW_3 \times \cW_3$), the dual ``IR $\cW_3$ CFT'' is deformed by irrelevant operators while the dual ``UV $\cW_3^{(2)}$ CFT'' is deformed by relevant operators both dual to the chemical potentials. If it was the case, turning on $\mu,\bar \mu$ would break both asymptotic symmetry algebras. In this paper, we will see that this picture is not realized. Instead, we will show that for each $sl(2,\mathbb R)$ embedding or, equivalently, each choice of boundary conditions, the asymptotic symmetry algebra does not depend on $\mu,\bar \mu$, though the generators get modified. More concretely, the asymptotic symmetry algebra for the principal embedding is always $\cW_3 \times \cW_3$, while the asymptotic symmetry algebra for the diagonal embedding is $\cW_3^{(2)}\times \cW_3^{(2)}$.  We are then led to conjecture that turning on a  chemical potential preserves the symmetries of the dual CFT (the conformal generators will however be modified). This conjecture is consistent with the fact that the gravity side \cite{Kraus:2011ds} agrees with the CFT calculation based on $\cW$ symmetry \cite {Gaberdiel:2012yb} at very high temperature.

Another view on higher spin black holes comes from the perspective of integrable systems. The phase space described by  Dirichlet boundary conditions at finite chemical potential is in fact described by the third equation in the KdV hierarchy known as the (good) Boussinesq equation \cite{Boussinesq:1872,Zacharov:1974,McKean:1981aa,Deift:1982}. It has been known since the early 90s that the Boussinesq system enjoys a bi-Hamiltonian structure both in standard evolution and evolution in the reverse coordinates (which are here the boundary lightcone coordinates $x^\pm$) \cite{Mathieu:1988pm,Mathieu:1991et}. The second Poisson structures coincide with the $\cW_3$ and $\cW_3^{(2)}$ algebras in standard and reverse evolution, respectively.  In this paper, we will show that there also exists a bi-Hamiltonian structure in $t=(x^++x^-)/2$ evolution defined with four functional initial data (which correspond to the boundary conditions for the diagonal embedding). The Poisson bracket defined from the second Hamiltonian structure will be shown to be isomorphic to the $\cW_3^{(2)}$ algebra. We will also derive the infinite tower of conserved charges in time evolution from the third KdV hierarchy. We will see that the charges differ from the standard charges (defined in $x^-$ evolution) only from a term linear in the chemical potentials.

The layout of this paper is the following. In section 2 we review the $sl(3,\mathbb R) \times sl(3,\mathbb R)$ Chern-Simons theory using the two $sl(2,\mathbb R)$ embeddings.  In section 3, we define Dirichlet boundary conditions with non-zero chemical potentials for each embedding. In section 4, we discuss the bulk equations of motion, relate them to integrable systems and review some results in the integrability literature. In section 5, we derive the asymptotic symmetry algebra $\cW_3 \times \cW_3$ in the principal embedding using canonical methods by doing perturbations in $\mu$. We also discuss the tower of KdV charges. In section 6, we derive the asymptotic symmetry algebra $\cW^{(2)}_3\times \cW^{(2)}_3$ for the diagonal embedding using both canonical and integrability methods.

\section{$SL(3,\mathbb R) \times SL(3,\mathbb R)$ Chern-Simons theory}

We consider the $3d$ pure higher spin theory in the Chern-Simons formulation with gauge group $SL(3,\mathbb{R})_L\times SL(3,\mathbb{R})_R$. The action reads as
\bea
S[A,\bar{A}]=S_{k}[A]+S_{-k}[\bar{A}]\label{SE}
\eea
where
\bea
S_{k}[A]&=&{k\over4\pi}\int_{\cM}  \tr (A \wedge dA+{2\over3} A\wedge A\wedge A).
\eea
The equations of motion are given by \be F \equiv dA+A \wedge A=0,\quad  \bar F \equiv d\bar{A}+\bar{A} \wedge \bar{A}=0.\ee

There are only two $sl(2,\mathbb R)$ embeddings into the $sl(3,\mathbb R)$ algebra, namely the principal and diagonal embedding. Each embedding allows to define a vielbein and spin connection and therefore a geometry, as well as additional fields, by branching the adjoint representation of $sl(3,\mathbb R)$ into $sl(2,\mathbb R)$ representations.

Chern-Simons theory in the principal embedding consists of the spin 2 field coupled to a spin 3 field while it consists of  the spin 2 field coupled to two spin 3/2 fields and one spin 1 field  in the diagonal embedding. Our conventions for the $sl(3,\mathbb R)$ generators as well as the field redefinition relating the two $sl(2,\mathbb R)$ embeddings can be read in Appendix \ref{conv}.

\subsection{Principal embedding}

Let us review some key properties of the principal embedding. We choose length units such that the $AdS_3$ vacuum has radius $l$. The corresponding Einstein theory then has Newton's constant $G = \frac{l}{4k}$. The asymptotic symmetry algebra for Dirichlet boundary conditions around that vacuum  was obtained in \cite{Bilal:1991cf,Bilal:1991dk} and rederived in \cite{Campoleoni:2010zq} (see also \cite{Henneaux:2010xg}). It is the (classical) Zamolodchikov algebra $\cW_3$ \cite{Zamolodchikov:1985aa} with Virasoro central charge
\bea
c = 6 k = \frac{3l}{2G}.
\eea
The most general solution satisfying the boundary conditions can be written in terms of two functions $\cL$, $\cW$ (which are interpreted as the vev of the right-moving $(2,0)$ stress-tensor and $(3,0)$ spin 3 current) and their bar analogue as
\bea
A &=& e^{-\rho L_0} (L_1 - \frac{1}{k}\cL(x^+) L_{-1} - \frac{1}{4 k} \cW(x^+) W_{-2}) dx^+  e^{\rho L_0} + L_0 d\rho ,\nn\\
\bar A &=& - e^{\rho L_0} (L_{-1} - \frac{1}{k}\bar \cL(x^-) L_{1} - \frac{1}{4 k}\bar  \cW(x^-) W_{2}) dx^-  e^{-\rho L_0} - L_0 d\rho .\label{AdS3}
\eea

Black hole solutions with spin 3 charges were obtained in \cite{Gutperle:2011kf} as
\bea
A&=&b^{-1} a(x^+,x^-) b+b^{-1}db,\quad \bar{A}= b\bar{a}(x^+,x^-) b^{-1}+bdb^{-1},\quad
b=e^{\rho L_0},\nn\\
a&=&\Big(L_1-{1\over k}\mathcal{L} L_{-1}-{1\over4k}\mathcal{W}W_{-2}
\Big)dx^+ \nn\\
&&+\mu\Big(W_2-{2\mathcal{L}\over k}W_0+{\mathcal{L}^2\over k^2}W_{-2}+{2 \mathcal{W}\over k}L_{-1}
\Big)dx^- ,\label{hsbh} \\
\bar a&=&-\Big(L_{-1}-{1\over k}\bar{\mathcal{L}}L_1-{1\over 4k}\bar{\mathcal{W}} W_2\Big)dx^-\nn\\
&&-\bar{\mu}\Big(W_{-2}-{2\bar{\mathcal{L}}\over k}W_0+{\bar{\mathcal{L}}^2\over k^2}W_{2}+{2\bar{\mathcal{W}}\over k}L_{1}
\Big)dx^+, \nn
\eea
where the solution is written in wormhole gauge and $\mu,\mathcal{L},\mathcal{W}$ and $\bar{\mu},\bar{\mathcal{L}},\bar{\mathcal{W}}$ are constants.  It has been shown that a higher spin gauge transformation exists such that the corresponding transformed metric admits a horizon \cite{Ammon:2011nk}. There exists four different branches of solutions which have a trivial holonomy around the thermal Euclidean circle \cite{David:2012iu}, which has been proposed as the criterium to define a higher spin black hole \cite{Gutperle:2011kf}. The solutions necessarily have a spin 3 chemical potential $\mu$ when the spin 3 charge is turned on.
At $\mu=\bar\mu=\mathcal{W}=\bar{\mathcal{W}}=0,$ the solution is just a BTZ black hole. At finite $\mu$ and $\bar{\mu},$ the solution (\ref{hsbh}) grows as $e^{4\rho}$ and therefore violates the asymptotic $AdS_3$ boundary conditions \eqref{AdS3}.

\subsection{Diagonal embedding}
The $AdS^{(2)}_3$ vacuum of the diagonal embedding has radius $l/2$ and the corresponding Einstein theory has Newton's constant $G = l/2k$. The asymptotic symmetry algebra for Dirichlet boundary conditions around that vacuum is the Polyakov-Bershadsky algebra $\cW^{(2)}_3$   \cite{Polyakov:1989dm,Bershadsky:1990bg} with Virasoro central charge and $U(1)$ Kac-Moody level (see \cite{Bilal:1992ub,deBoer:1993iz,Ammon:2011nk,Castro:2012bc})
\bea
\hat c = \frac{3k}{2} = \frac{c}{4},\qquad \hat k=- {k\over3}.
\eea

The most general solution satisfying the boundary conditions can be written in terms of four functions $\cT$, $\cG^\pm$, $\cJ$ (which are interpretated as the vev of the left-moving $(2,0)$ stress-tensor, two bosonic $(3/2,0)$ fields and a $(1,0)$ current) and their bar analogue as
\bea
A &=& e^{-\rho \hat L_0}(  \hat L_{1}+\frac{6}{k} \cJ (x^-) \hat J_0 +\frac{4}{k} (\mathcal G^-  (x^-)\hat G^-_{-1/2}+\mathcal G^+  (x^-)\hat G^+_{-1/2})\nn\\
&&+\frac{4}{k} (\cT (x^-)+\frac{3}{k}\cJ^2 (x^-)) \hat L_{-1}   )\,   dx^-  e^{\rho \hat L_0}+ \hat L_0 d\rho ,\nn \\
\bar A &=& -e^{\rho \hat L_0}(  \hat L_{-1}+\frac{6}{k} \bar\cJ (x^+) \hat J_0 +\frac{4}{k} (\bar\cG^-  (x^+)\hat G^-_{1/2}+\bar\cG^+  (x^+)\hat G^+_{1/2})\nn\\
&&+\frac{4}{k} (\bar\cT (x^+)+\frac{3}{k}\bar\cJ^2 (x^+)) \hat L_{1}   )\,   dx^-  e^{-\rho \hat L_0}- \hat L_0 d\rho. \label{AdS32}
\eea

One can also rewrite the black hole solution in a form explicit for the diagonal embedding as
\bea
a &=& \sqrt{2}\lambda \Big(  \hat G_{1/2}^-  -\hat G_{1/2}^+ +\frac{6}{k}\cJ (\hat G^+_{-1/2}+\hat G^-_{-1/2} ) +\frac{2}{ k} (\cG^++\cG^-)\hat L_{-1} \Big) dx^+ \nn\\
&&\hspace{-1cm} + \Big( \hat L_{+1}+\frac{6}{k} \cJ  \hat J_0 +\frac{4}{k} (\mathcal G^- \hat G^-_{-1/2}+\mathcal G^+ \hat G^+_{-1/2})+\frac{4}{k} (\cT+\frac{3}{k}\cJ^2) \hat L_{-1}  \Big) dx^-\label{bhw32}
\eea
and similar expressions for the barred sector. The hatted $sl(3,\mathbb R)$ generators are related to the ones without hat by linear combinations. For the black hole solutions, $\mathcal{J},\mathcal{G}^\pm,\mathcal{T}$ are also constants, and are related to $\mathcal{L}$ and $\mathcal{W}$ by
\bea
\mathcal G^\pm &=& {\sqrt{2}}\mu^{3/2}  \cW ,\nn \\
\cJ  &=& -\frac{2}{3}\mu \cL ,\label{fr0} \\
\cT &=& -\frac{16\mu^2}{3k} \cL^2. \nn
\eea
The chemical potential
\bea
\lambda =\frac{1}{2\sqrt{\mu}}
\eea
is then recognized as a spin 3/2 chemical potential. The solution violates the boundary conditions for $AdS^{(2)}_3$ \eqref{AdS32}.

\subsection{Embedding versus boundary conditions}

In this subsection, we make connections between different choices of embedding and different choices of boundary conditions. The canonical analysis of boundary conditions can always be formulated as an initial data problem. Given a Cauchy surface, boundary conditions are given by  some fall-off conditions on the initial data. For Chern-Simons theory around an $AdS_3$ background, a gauge choice has separated out the radial dependence in the reduced connections $a,\,\bar a$ in both choices of embedding.  What makes the two choices of embedding distinct is the different choices of initial data. We can see explicitly that in the principal embedding (\ref{AdS3}), there are two initial data, $\mathcal{L}(0,\phi)$ and $\mathcal{W}(0,\phi)$ while in the diagonal embedding (\ref{AdS32}), there are four initial data,  $\cT(0,\phi)$, $\cG^\pm(0,\phi)$, $\cJ(0,\phi)$.

In the following, we will discuss consistent boundary conditions in each embedding that include the black holes (\ref{hsbh}) or (\ref{bhw32}). Our boundary conditions are natural generalizations of the Dirichlet boundary conditions (\ref{AdS3}) or (\ref{AdS32}) at finite chemical potentials.

\section{Dirichlet boundary conditions at finite $\mu$}
\subsection{Principal embedding}
We work in radial gauge where $A_\rho = 1,\; \bar A_\rho = -1$.
The connection can be expressed in terms of the reduced connections $a,\bar a$ as
\bea
A &=& b^{-1}a(x^+,x^-)b+b^{-1}db,\qquad \bar A = b \bar a(x^+,x^-)b^{-1}+bdb^{-1},\nn\\
b(\rho) &=& e^{\rho L_0}.\nn
\eea
The Dirichlet boundary conditions at $\rho \rightarrow \infty$ which generalize the Brown-Henneaux boundary conditions in the presence of fixed constant spin 3 chemical potentials $\mu$, $\bar{\mu}$ can be expressed in terms of fall-off conditions together with a specification of the initial data problem.

First, at fixed $\mu$, the fall-off conditions can be expressed as
\bea
a_+ &=& L_1 - \frac{1}{k} \cL(x^+,x^-) L_{-1}-\frac{1}{4 k}  \cW(x^+,x^-) W_{-2},\nn\\
a_- &=& \mu W_2 + (\text{higher}),\nn\\
\bar a_- &=& - \left( L_{-1} - \frac{1}{k} \bar \cL(x^+,x^-) L_{1}-\frac{1}{4 k}  \bar \cW(x^+,x^-) W_{2}\right) ,\nn\\
\bar a_+  &=& -\bar \mu W_{-2} + (\text{lower}),\label{dbc}
\eea
where (higher) (resp. (lower)) are terms linear in higher (resp. lower) weight $sl(3,\mathbb R)$ generators which correspond to terms that fall-off quicker at infinity in $A$, $\bar A$.

Second, in the principal embedding, we require that the initial data at $t=0$ be entirely specified using the values of $\mathcal{L}$ and $\mathcal{W}$ at $t=0$: $\cL(0,\phi)$, $\cW(0,\phi)$.

To summarize, the Dirichlet boundary conditions for the principal embedding consist of the fall-off conditions given in (\ref{dbc}), with $\cL(0,\phi)$, $\cW(0,\phi)$ as the initial data.

\subsection{Diagonal embedding}
 A second, equivalent, way of stating the fall-off conditions is to impose
\bea
a_- &=&  \hat L_{+1}+\frac{6}{k} \cJ  \hat J_0 +\frac{4}{k} (\mathcal G^- \hat G^-_{-1/2}+\mathcal G^+ \hat G^+_{-1/2})+\frac{4}{k} (\cT+\frac{3}{k}\cJ^2) \hat L_{-1},\nn\\
a_+ &=& \sqrt{2}\lambda ( \hat G_{1/2}^-  -\hat G_{1/2}^+) + (\text{lower}),\nn\\
\bar a_+ &=& -\left( \hat L_{-1}+\frac{6}{k} \bar \cJ  \hat J_0 +\frac{4}{k} (\bar \cG^- \hat G^-_{1/2}+\bar \cG^+ \hat G^+_{1/2})+\frac{4}{k} (\bar \cT+\frac{3}{k}\bar \cJ^2) \hat L_{1}\right),\nn\\
\bar a_- &=& - \sqrt{2}\lambda ( \hat G_{-1/2}^-  - \hat G_{-1/2}^+) + (\text{higher}),\label{dbc2}
\eea
where we recall that the hatted $sl(3,\mathbb R)$ generators are defined in \eqref{fieldr}.

The two fall-off conditions \eqref{dbc} and \eqref{dbc2} are equivalent because the field equations $F=\bar F =0$ completely fix the form of the gauge field. They impose
\bea
a&=&\Big(L_1-{1\over k}\mathcal{L} L_{-1}-{1\over4k}\mathcal{W}W_{-2}
\Big)dx^+ \nn\\
&&\hspace{-2cm}+\mu\Big(W_2-{2 \mathcal{L}\over k}W_0+{2\over3k}\p_+\mathcal{L}W_{-1} +({\mathcal{L}^2\over k^2}-{1\over6k}\p_+^2\mathcal{L})W_{-2}+{2 \mathcal{W}\over k}L_{-1}
\Big)dx^- ,\label{ncbc}
\eea
or equivalently,
\bea
a &=& \sqrt{2}\lambda \Big(  \hat G_{1/2}^-  -\hat G_{1/2}^+ +\frac{6}{k}\cJ (\hat G^+_{-1/2}+\hat G^-_{-1/2} ) +\frac{2}{ k} (\cG^++\cG^-)\hat L_{-1} \Big) dx^+ \nn\\
&&\hspace{-1cm} + \Big( \hat L_{+1}+\frac{6}{k} \cJ  \hat J_0 +\frac{4}{k} (\mathcal G^- \hat G^-_{-1/2}+\mathcal G^+ \hat G^+_{-1/2})+\frac{4}{k} (\cT+\frac{3}{k}\cJ^2) \hat L_{-1}  \Big) dx^-,\label{ncbc2}
\eea
and similarly for the bar connection. In passing from the $AdS_3$ formulation \eqref{ncbc} to the $AdS_3^{(2)}$ formulation \eqref{ncbc2}, we used the shift of radius  $\rho = \frac{\hat \rho}{2} + \Lambda$, which corresponds to the following gauge transformation
\bea
a \rightarrow e^{-\Lambda L_0} \, a\,  e^{\Lambda L_0},\qquad e^\Lambda \equiv \lambda = \frac{1}{2\sqrt{\mu}}
\eea
in order to normalize the coefficient of $\hat L_1$ in $a_-$ to 1. We also used the field redefinition
\bea
\mathcal G^\pm &=& \frac{\sqrt{2}}{3}\mu^{3/2} (3 \cW \pm \p_+ \cL),\nn \\
\cJ  &=& -\frac{2}{3}\mu \cL ,\label{fr} \\
\cT &=& \frac{2\mu^2}{3}(\p_+^2 \cL - \frac{8}{k}\cL^2). \nn
\eea
or, conversely,
\bea
\cL = -6 \lambda^2 \cJ,\qquad \cW = 2\sqrt{2}\lambda^3 (\cG^++\cG^-).\label{frtwo}
\eea

The Dirichlet boundary condition for the diagonal embedding consists of the fall-off conditions (\ref{ncbc2}), together with the specification of $\mathcal{T},\,\mathcal{J},\,\mathcal{G}^\pm$ at $t=0$ or, equivalently, $\mathcal{L}(0,\phi),\,\mathcal{W}(0,\phi),\,\dot{\mathcal{L}}(0,\phi),\,\ddot{\mathcal{L}}(0,\phi) $ as the initial data, where the dot denotes a derivative with respect to $t$.

\subsection{Variational principle}

The variation of the bulk Chern-Simons action is non-zero,
\bea \delta S_k[A]&=&-{k\over4\pi}\int dx^+dx^- \hbox{Tr} (a_+\delta a_--a_-\delta a_+),\\
\delta S_{-k}[\bar{A}]&=&{k\over4\pi}\int dx^+dx^- \hbox{Tr} (\bar{a}_+\delta \bar{a}_--\bar{a}_-\delta \bar{a}_+).
\eea
After adding the boundary terms found by \cite{deBoer:2013gz}
\bea
I_{Bdy}&=&-{k\over2\pi}\int_{\p M} d^2x \hbox{Tr}[(a_+-2L_1)a_-]-{k\over2\pi}\int _{\p M} d^2 x \hbox{Tr}[(\bar{a}_++2\bar{L}_{-1})\bar{a}_+]\nn\\
&=&-{1\over2\pi}\int_{\p M} d^2x \big( \mu \mathcal{W}+\bar{\mu}\bar{\mathcal{W}}\big),
\eea
the variation of the action becomes \be \delta S={1\over\pi}\int_{\p M} d^2x\big( \mathcal{W}\delta\mu+\bar{\mathcal{W}}\delta\bar{\mu}\big)=0,\ee
since we hold $\mu$ and $\bar \mu$ fixed. Therefore we see that the fall-off conditions (\ref{dbc}) or equivalently \eqref{dbc2} lead to good variational principle.

\section{Equations of motion}
\label{sec:EOM}

Starting from the boundary conditions (\ref{dbc})-\eqref{dbc2}, the equations of motion completely fix the form of the connections as  \eqref{ncbc}-\eqref{ncbc2}. The remaining equations of motion reduce to the following set of coupled partial differential equations
 \bea
\p_-\mathcal{L}&=&-2\mu \p_+\mathcal{W}, \nn\\
\p_-\mathcal{W}&=&{2\mu \over3 }\p_+ \Big( \p_+^2\mathcal{L}-\frac{8}{k} \mathcal{L}^2\Big),\label{sol}
\eea
or, equivalently, to
\bea
\p_+ \cJ &=& -\sqrt{2} \lambda  (\cG^+ - \cG^-), \nn\\
\p_+ \cT &=& \frac{\sqrt{2}}{2 }\lambda (\p_- \cG^+ + \p_- \cG^-),\nn\\
\p_+ \cG^\pm &=& \sqrt{2} \lambda (\frac{3}{2} \p_- \cJ \pm (\cT + \frac{12}{k}\cJ^2)),\label{sol2}
\eea
after using the relations \eqref{fr}-\eqref{frtwo}. Similar equations hold for the bar sector with $x^\pm$ interchanged.

When $\mu = 0$,  the phase space in the principal $sl(2,\mathbb R)$ embedding is clear.
 The bulk equations of motion determine that fields are right-moving $\cL = \cL(x^+)$, $\cW = \cW(x^+)$. The initial data is indeed given by $\cL,\cW$ at $t=0$.  For the diagonal embedding, (\ref{sol2}) seems singular, but this is just due to the singular rescaling (\ref{fr}).
 After proper rescaling by factors of $\mu$, the new fields $\mu^{-1}\mathcal{J},\mu^{-3/2}\mathcal{T},\mu^{-2}\mathcal{G}^\pm $ will also become purely right moving and their value at $t=0$ specifies the initial data.

When $\mu = \infty$ ($\lambda = 0$), the defining functions $\cJ,\cG^\pm,\cT$ in the diagonal embedding are left-moving, $\cJ(x^-),\cG^\pm(x^-),\cT(x^-)$. Using the field redefinition,
the properly rescaled fields $\mu\cL$ and $\mu^{3/2}\cW$ will also be left-moving.

For generic finite values of $\mu$, the functions have a non-trivial $x^\pm$ dependence, to which we now turn our attention.


\subsection{Boussinesq system on the light-cone}

Let us make a connection between the equations of motion and integrable systems. First, the equations of motion \eqref{sol} are precisely the Boussinesq equations with light-cone coordinate $x^-$ as formal time evolution coordinate. Following the conventions of Mathieu and Oevel \cite{Mathieu:1991et}, we set
\bea
u &\equiv & \frac{12\lambda^2}{k}\cJ = -\frac{2}{k}\cL, \nn\\
v &\equiv & \frac{4\sqrt{6}}{k}\lambda^3 (\cG^+ +\cG^-) = \frac{2\sqrt{3}}{k}\cW ,\label{defuv}
\eea
and we define the rescaled light-cone coordinate
\bea
\hat x^- & \equiv &  \frac{2}{\sqrt{3}}\mu x^-.
\eea
The system \eqref{sol} becomes the Boussinesq equation ($Bsq$ for short)
\bea
\p_{\hat x^-} \left( \begin{array}{c} u \\ v \end{array} \right) =  \left( \begin{array}{c} \p_+ v \\ -\p_{+}^3 u -8 u \p_+ u \end{array} \right) ,
\eea
which can be formulated as
\bea
\p_{\hat x^-}^2 u = \p_+^4 u + 4\p_+^2 (u^2). \label{Bsq}
\eea
More precisely, there are two classes of Boussinesq equations which differ by a sign. The equation \eqref{Bsq} is known as the ``good'' Boussinesq equation \cite{McKean:1981aa}. It can also be recognized as the reduction of the $2+1$ Kadomtsev--Petviashvili (KP) equation to time independent solutions (with the two spatial directions played here by $x^+,\hat x^-$). There are similarly two distinct versions of the KP equation, referred to as KPI and KPII, differing by a sign. The equation \eqref{Bsq} is the KPII equation for time-independent solutions.

The inversion of the light-cone coordinates $\hat x^-,\, x^+$ leads to the Boussinesq equation in inverted variables ($\widetilde{Bsq}$ for short)
\bea
\p_{x^+} \left( \begin{array}{c} u \\ v \\ w \\ z \end{array} \right) =  \left( \begin{array}{c} w \\ \p_{\hat x^-} u \\ z-4u^2 \\-\p_{\hat x^- }v \end{array} \right) .
\eea
Our second observation is that these equations are precisely the equations of motion \eqref{sol2} after the field redefinition \eqref{defuv} accompanied by
\bea
w &\equiv & -\frac{12\sqrt{2}\lambda^3}{k}(\cG^+ - \cG^-),\nn \\
z &\equiv & -\frac{48\lambda^4}{k}\cT.
\eea

Both the Boussinesq $Bsq$ and $\widetilde{Bsq}$ equations are integrable systems with a bi-Hamiltonian structure. Let us already note however that these Hamiltonian structures are defined at constant $\hat x^-$ and $x^+$, respectively. Here, we have $x^\pm = t \pm \phi$ with $\phi \sim \phi +2\pi$ and the Cauchy evolution is along $t$. We cannot therefore directly use these structures on the constant $t$ slice in order to build the symmetry algebra of conserved charges.
We will explicitly construct  the bi-Hamiltonian structure of our system on the constant $t$ slice in Section \ref{biH}, which is a new result to our knowledge. In this section, we will continue to explore the known structure of the Boussinesq equations by reviewing the bi-Hamiltonian structure and the infinite tower of commuting charges. The reader familar with this material might skip the remainder of this section and jump to Section \ref{prin}.

\subsection{Standard bi-Hamiltonian structures}
\label{biH}

Let us quickly review the Hamiltonian structures for the $Bsq$ equations (in $x^-$ evolution) and $\widetilde{Bsq}$ equations (in $x^+$ evolution), which  encode in an elegant algebraic way the $\cW_3$ and $\cW_3^{(2)}$ algebras.

 The two Hamiltonians for the $Bsq$ equation are given by
\bea
\p_{\hat x^-} \left( \begin{array}{c} u \\ v \end{array} \right) = \Theta^1 \left( \begin{array}{c} \delta/\delta u \\ \delta / \delta v \end{array} \right) \cH_1 = \Theta^2 \left( \begin{array}{c} \delta/\delta u \\ \delta / \delta v \end{array} \right) \cH_2,
\eea
\bea
\cH_1 = \int dx^+ \left( \frac{1}{2} u_{x^+}^2 - \frac{4}{3}u^3+\frac{1}{2}v^2 \right),\qquad \cH_2 = \int dx^+ \frac{1}{2} v,
\eea
where the two Hamiltonian operators $\Theta^{1}$, $\Theta^{2}$ are given by
\begin{small}
\bea
\Theta^{1}\hspace{-6pt} &=&\hspace{-6pt} \left( \begin{array}{cc} 0&\p \\ \p & 0
\end{array}\right) ,\label{th21}\\
\Theta^{2}\hspace{-6pt} &=&\hspace{-6pt} \left( \begin{array}{cc} \p^3 + 2u\p + u_x & 3 v \p +2 v_x \\
3 v \p +v_x & -(\p^5 +10 u \p^3+15 u_x \p^2 +(9 u_{xx}+16 u^2)\p +2 u_{xxx}+16 u u_x)
\end{array}\right) ,\nn
\eea
\end{small}
\hspace{-6pt}and where we dropped the superscript index $x=x^+$, $\p = \p_+$ to shorten the notation. The second Hamiltonian structure defines a Poisson bracket among the fields $(u_1=u,u_2=v)$,
\bea
\{ u_i(x^+) , u_j (y^+) \} = (\Theta^{2})_{ij} \delta(x^+-y^+),\qquad i=1,2,
\eea
at fixed $x^-$. If $x^+$ was a periodic coordinate, one could Fourier decompose in modes along $x^+$ and the Poisson bracket would then exactly correspond to the $\cW_3$ algebra.

The two Hamiltonians for the $\widetilde{Bsq}$ equation are given by
\bea
\p_{\hat x^+} \left( \begin{array}{c} u \\ v \\ w \\ z \end{array} \right) = \Theta^1 \left( \begin{array}{c} \delta/\delta u \\ \delta / \delta v \\ \delta / \delta w \\ \delta /\delta z \end{array} \right) \cH_1 = \Theta^2  \left( \begin{array}{c} \delta/\delta u \\ \delta / \delta v \\ \delta / \delta w \\ \delta /\delta z \end{array} \right)\cH_2,
\eea
\bea
\cH_1 = \int d\hat x^- \left(\frac{4}{3}u^3 + \frac{1}{2}(v^2+w^2)- u z \right),\qquad \cH_2 = \int d\hat x^- \, v.
\eea
The two Hamiltonian operators $\Theta^{1}$, $\Theta^{2}$ are given by
\bea
\hspace{-15pt}{\Theta}^{1} \hspace{-6pt}&=&\hspace{-6pt}  \left(
\begin{array}{cccc}
0&0&1&0\\
0&0&0&-\p_x\\
-1&0&0&0\\
0&-\p_x &0&0
 \end{array}
\right) ,\label{th12} \\
\hspace{-15pt}{ \Theta}^{2} \hspace{-6pt}&=& \hspace{-6pt} \left(
\begin{array}{cccc}
-\p_x &  w & 3  v & -2  u_x-2 u \p_x \\
- w &  u_x+2 u \p_x & \p_x^2 - z+4 u^2 & -2  v_x -3 v\p_x \\
-3  v &-\p_x^2 +  z - 4 u^2 & -3 u_x - 6  u\p_x & -2 w_x - 3 w\p_x \\
-2 u \p_x& - v_x -3  v \p_x & -  w_x - 3  w \p_x &\p_x^3 -2 z_x - 4 z \p_x
 \end{array}
\right),\label{th22}
\eea
where we wrote $x = \hat x^-$ here in order to shorten the notation.  If $\hat x^-$ was a periodic coordinate, one could Fourier decompose in modes along $\hat x^-$ and the Poisson bracket would then exactly correspond to the $\cW^{(2)}_3$ algebra.

\subsection{Commuting charges from the KdV hierarchy}

The Boussinesq equation is the first non-trivial field equation from the $n=3$ KdV hierarchy (for reviews, see e.g. \cite{Batlle:1992uu,Dickey:1997aa}). An infinite set of mutually commuting conserved charges can be obtained from the $n=3$ KdV hierarchy. These charges are however defined on a constant $x^-$ slice and are integrated along $x^+$, which is unsuitable for our problem.
We will make connection with canonical methods and define charges on the constant $t$ slice in Section \ref{sec:can}. Here, we proceed with our review.

We can reformulate the field equations \eqref{sol} in the language of the KdV hierarchy for level $n=3$ as follows. We introduce the level $n=3$ Lax operator
\bea
L =  \p_+^3 - \frac{4}{k} \cL \p_+ - \frac{2}{k} (\p_+ \cL - \cW) \label{Laxo}
\eea
which acts on the space of functions of $x^\pm$. It is natural to consider the Gel'fand-Dickey ring of pseudo-differential operators generated by $\p_+^k$ where $k \in \mathbb Z$ \cite{Gelfand:1987aa}. One can then take a fractional power of the operator $L$ as
\bea
L^{1/3} &=&  \p_+ - \frac{4}{3k}\cL \p_+^{-1}+O(\p_+^{-2}) ,\\
L^{2/3} &=&  \p_+^2 - \frac{8}{3k}\cL +O(\p_+^{-1}).
\eea
The equations of motion \eqref{sol} are then recognized as the Lax equations
\bea
(2\mu)^{-1} \p_- L = [L^{2/3}_+,L],
\eea
where the subindex $\mbox{}_+$ indicates the truncation to non-negative powers of $\p_+$ only.  The factor of $\mu$ can be absorbed into a redefinition of $x^-$. The $n=3$ KdV hierarchy can be written as
\bea
(2\mu)^{-1} \p_- L^{k/3} =  [L_+^{2/3}, L^{k/3}]
\eea
for any integer $k \geq 1$. All conserved quantities of the Boussinesq equation in $x^-$ evolution can be expressed as
\bea
\mathcal C_k = \int dx^+ h_k ,\qquad h_k \equiv \text{Res} ( L^{k/3}),\label{Ck}
\eea
for any positive integer $k \geq 1$ and $k \neq 3 \mathbb N$, where $\text{Res}(\cdot )$ denotes the residue or term proportional to $\p_+^{-1}$ in the argument. Remark that $\p_+$-exact terms in $h_k$ do not contribute to the conserved charges. The conservation of these quantities follows from the property
\bea
(2\mu)^{-1}  \p_-h_k= \p_+ l_k \label{cons}
\eea
where $l_k$ can be constructed from the hierarchy as
\bea
\text{Res} [L_+^{2/3}, L^{k/3}] = \p_+ l_k.\label{lk}
\eea
By direct evaluation, the first four charge densities in the $n=3$ KdV hierarchy are explicitly given by
\bea
h_1 &=& -\frac{4}{3k} \cL,\\
h_2 &=& \frac{4}{3k}\mathcal W,\\
h_4&=& -\frac{32}{9k^2}\cL \cW+\p_+(\cdot ),\\
h_5 &=& \frac{20}{9k^2}(\cW^2+\frac{1}{3} (\p_+\cL)^2+\frac{16}{9k}\cL^3 )+\p_+(\cdot ).\label{hk}
\eea

\section{$\cW_3$ symmetry in the Principal embedding}\label{prin}

In this section, we will first derive the canonical infinitesimal charges and review the $\cW_3$ asymptotic symmetry at $\mu = 0$. We will then show that it exists a basis of symmetry generators which preserves the $\cW_3$ symmetry when $\mu$ is turned on perturbatively. We will finally build commuting charges from canonical methods at finite $\mu$ and make a connection with a linear deformation in $\mu$ of the
integrable tower of commuting charges from the Boussinesq hierarchy.

\subsection{Infinitesimal symmetries}

Asymptotic symmetries can be identified with the gauge transformations
\bea
\delta A_\mu &=& D_\mu \Lambda = \p_\mu \Lambda + [A_\mu,\Lambda]
\eea
which preserve the phase space (\ref{ncbc}). They are given by
\bea
 \Lambda &=& e^{-\rho L_0}\lambda e^{\rho L_0},\qquad \lambda=\sum_{i=-1}^{1} \epsilon^{(i)}L_i+\sum_{m=-2}^2 \chi^{(m)}W_m
\eea
where
\be
\epsilon \equiv \epsilon^{(1)}(x^+,x^-), \qquad \chi\equiv \chi^{(2)} (x^+,x^-),
\ee
obey the following system,
\bea
\p_-\chi&=&2\mu\p_+\epsilon,\nn\\
\p_-\epsilon&=&- {2\mu\over3}\p_+^3\chi+{32\mu\over3k}\mathcal{L}\p_+\chi\, .
\label{epchi}
\eea
The remaining components $\eps^{(0),(-1)}$, $\chi^{(1),(0),(-1),(-2)}$ are auxiliary functions which are fixed in terms of $\epsilon$, $\chi$ and the fields $\cL,\cW$. Under a gauge transformation,
the fields $\mathcal{L}$ and $\mathcal{W}$ transform as
\bea
\delta \mathcal{L}&=&-\p_+\mathcal{L}\epsilon-2\mathcal{L}\p_+\epsilon+{k\over2}\p_+^3\epsilon+2\chi\p_+\mathcal{W}+3\p_+\chi\mathcal{W},\nn \\
\delta\mathcal{W}&=&-\epsilon\p_+\mathcal{W}-3\p_+\epsilon\mathcal{W}\nn\\
&&-{1\over3}\Big(2\chi\p_+^3\mathcal{L}+9\p_+\chi\p_+^2\mathcal{L}+15\p_+^2\chi\p_+\mathcal{L}+10\p_+^3\chi\mathcal{L}-{k\over2}\p_+^5\chi
\nn\\&&-{32\over k}(\chi\mathcal{L}\p_+\mathcal{L}+\p_+\chi\mathcal{L}^2)\Big).\label{trsf}
\eea
These transformation laws can be expressed in terms of the Poisson bracket for the second Hamiltonian structure of $\cW_3$ \eqref{th21}. They are independent of $\mu$. Nevertheless, it does not imply that $\cW_3$ is the asymptotic symmetry algebra at finite $\mu \neq 0$ since the conserved charges are not proportional to $\cL$ and $\cW$ when $\mu \neq 0$ as we will see shortly.

For the $\bar A$ sector, gauge transformations
\bea
\delta \bar A_\mu = \bar D_\mu \bar \Lambda = \p_\mu \bar \Lambda + [\bar A_\mu,\bar \Lambda]
\eea
preserving the boundary conditions are given by
\be
\bar{\Lambda}=e^{\rho L_0}\bar{\lambda} e^{-\rho L_0},\qquad \bar{\lambda}=-\sum_{i=-1}^{1} \bar{\epsilon}^{(i)}L_i-\sum_{m=-2}^2 \bar{\chi}^{(m)}W_m
\ee
where
\be
\bar \epsilon \equiv\bar  \epsilon^{(-1)}(x^+,x^-), \qquad \bar \chi\equiv \bar \chi^{(-2)} (x^+,x^-),
\ee
obey the following system,
\bea
\p_+\bar{\chi}&=&2\bar{\mu}\p_-\bar{\epsilon}\\
\p_+ \bar{\epsilon}&=&- {2\bar{\mu}\over3}\p_-^3\bar{\chi}+{32\bar{\mu}\over3k}\bar{\mathcal{L}}\p_-\bar{\chi}
\label{epchibar}
\eea
and $\bar{\epsilon}^{(0),(1)}$, $\bar{\chi}^{(-1),(0),(1),(2)}$ are auxiliary dependent functions. The fields $\bar{\mathcal{L}}$ and $\bar{\mathcal{W}}$ transform exactly as \eqref{trsf} where all quantities are barred and $x^\pm$ are interchanged.


The infinitesimal conserved charges associated with the gauge parameters $\Lambda$ and $\bar\Lambda$ are given by
\be \sdelta Q_{\Lambda,\bar{\Lambda}}=\frac{k}{2\pi}\int_{\Sigma} dx^i\; \hbox{Tr} \left( \delta a_i \lambda - \delta \bar a_i \bar \lambda \right)\ee
where $\Sigma$ is a one-dimensional slice. The $\rho$ dependence completely factorizes so that the charges are defined at any value of $\rho$. For the $A$ sector,
\bea
\sdelta Q_{\epsilon,\chi}
&=&{1\over2\pi}\int_{\Sigma} (\epsilon\delta\mathcal{L}-\chi\delta\mathcal{W})dx^+
\label{chargephi} \\ \nn
&&-\Big(2\mu \epsilon\delta\mathcal{W}+{2\mu\over3}(-{16\over k}\chi\cL \delta\mathcal{L}-\p_+\chi\delta\p_+\mathcal{L}+\p_+^2\chi\delta\mathcal{L}+\chi\p_+^2\delta\mathcal{L})\Big)dx^-\eea
and a similar expression holds for the $\bar A$ sector. For a fixed $t$ slice, the charges cannot be explicitly integrated without knowing the general solution of \eqref{epchi} for $\eps$, $\chi$.

\subsection{Perturbation in $\mu$}

Let us first discuss the $A$ sector. When $\mu =0$, the gauge parameters are given by field-independent right-moving functions, $\eps = \eps(x^+)$, $\chi = \chi(x^+)$. The charges are integrable and given by
\bea
\cQ^{\mu = 0}_{(\eps,\chi)} = \int d\phi \left( \eps(x^+) \cL(x^+) -\chi(x^+) \cW(x^+) \right) .\label{W3c}
\eea
Using \eqref{trsf}, we can rederive that the charges represent the $\cW_3$ algebra under the canonical Poisson bracket.

Let us now obtain the algebra after we turn on a $\mu$ deformation. A priori, we do not know what the algebra is, since it is generally believed  that the $\cW_3 \times \cW_3$ symmetry is broken. To find the algebra in the principal embedding,
our strategy is to do perturbation theory around  $\mu = 0$.
At $\mu=0$, the boundary conditions only require two initial data $\mathcal{L}(0,\phi)$ and $\mathcal{W}(0,\phi)$. All time derivatives are determined by the equations of motion.
The symmetry preserved by the boundary conditions is also parameterized by two initial data $\epsilon(0,\phi),\chi(0,\phi)$.
 After turning on $\mu$, as we will see explicitly below, the equations of motion can be expanded to any given order in $\mu$, which express $\dot{\mathcal{L}}(0,\phi)$ and $\dot{\mathcal{W}}(0,\phi)$ in terms of $\mathcal{L},\mathcal{W}$ and their spatial derivatives. Therefore, the initial data problem with $\mathcal{L}(0,\phi),\mathcal{W}(0,\phi)$ as initial data is always well defined at any order in $\mu$. The equations of motion then imply that the infinitesimal charges are conserved in time. It is therefore sufficient to build the asymptotic symmetry algebra on the initial time slice. To get the algebra, we need to get the infinitesimal conserved charges associated with the symmetry in a good basis. Again, the basis at $\mu=0$ will be our starting point, namely $\Lambda_L\equiv (\epsilon,\chi)=(\tilde\epsilon(0,\phi),0)$ generates the Virasora algebra, while
$\Lambda_W\equiv (\epsilon,\chi)=(0,\tilde\chi(0,\phi))$ generates the spin 3 algebra, where $\tilde\epsilon(0,\phi),\tilde\chi(0,\phi)$ are the field-independent initial data for the gauge generators. After turning on $\mu$, we will determine the basis for $ (\epsilon,\chi)$ by two criteria: first, the associated infinitesimal charges should be integrable and second, the resulting algebra should be as close as possible to the $\cW_3$ algebra. After trial and error, it turns out that we can choose a basis such that the final algebra is still exactly $\cW_3$. We did the explicit calculation up to $\mathcal{O}(\mu^4)$ but we expect that this result will extend to all orders in perturbation theory.

Let us now obtain the $\cW_3$ algebra in perturbation theory around $\mu = 0$.  We performed the expansion up to $O(\mu^4)$ with the help of Mathematica\footnote{Our code can be obtained by request via email.}.
The equations of motion \eqref{sol} can be expanded as
\bea
\dot{\mathcal{L}}&=&\mathcal{L}'-4\mu \mathcal{W}' -\frac{8\mu^2}{3} \p_\phi \big(\mathcal{L}''-\frac{8}{k} \mathcal{L}^2\big)+16\mu^3\big(\mathcal{W}'''-{16\over3k}\mathcal{L}\mathcal{W}'\big)\nn\\
&&+ O(\mu^4),\label{eqLt}\\
\dot{\mathcal{W}}&=&\mathcal{W}'+{4\mu \over3 }\p_\phi \big( \p_\phi^2\mathcal{L}-\frac{8}{k} \mathcal{L}^2\big)-{8\mu^2}\big( \mathcal{W}'''- \frac{16}{3k} \mathcal{L}\mathcal{W}')\\
&&-{32\mu^3 \over9k}\Big(3\big(k\p_\phi^5\mathcal{L}-48\mathcal{L}'\mathcal{L}''\big)-56\mathcal{L}\mathcal{L}'''+{128\over k}\mathcal{L}^2\mathcal{L}'\Big)+ O(\mu^4) ,\nn
\eea
where dots denote time derivatives and primes $\phi$ derivatives.
Higher order time derivatives can be obtained by using the equations of motion successively.
The equations for the gauge parameters also become
\bea
\dot \eps &=& \eps' -\frac{4\mu}{3} \big( \chi''' - \frac{16}{k}\cL \chi' \big) -8\mu^2 \big( \eps''' - \frac{16}{3k} \cL \eps' \big)\\
&&+{32\mu^3 \over9k}\Big(3\big(k\p_\phi^5\mathcal{\chi}-32\mathcal{L}'\mathcal{\chi}''-16\mathcal{L}''\chi'\big)-56\mathcal{L}\mathcal{\chi}'''+{128\over k}\mathcal{L}^2\mathcal{\chi}'\Big)+O(\mu^4),\nn\\
\dot\chi &=& \chi'+4\mu \eps' -\frac{8\mu^2}{3}\big( \chi''' - \frac{16}{k}\cL \chi' \big) -16\mu^3\big(\epsilon'''-{16\over3k}\mathcal{L}\epsilon'\big)+O(\mu^4)\label{eqet} .
\eea
The initial data on a constant $t$ slice is therefore $\cL(0,\phi),\; \cW(0,\phi)$ for the fields and correspondingly $\eps(0,\phi)$, $\chi(0,\phi)$ for the gauge parameters. All  derivatives
with respect to time are determined by using the equations of motion \eqref{eqLt}-\eqref{eqet} up to the order $O(\mu^4)$. One can check that the infinitesimal charges \eqref{chargephi} are conserved in time after using \eqref{eqLt}-\eqref{eqet}. We can therefore concentrate our attention on the initial time slice.

After a large amount of trial and error, we take as an ansatz for the gauge parameters associated with the Virasoro generator,
\bea
\Lambda_L& \equiv &(\epsilon,\,\chi)\nn \\
&=&\big(1+{3\over2}\mu^2\p_\phi^2,  -\mu-\frac{1}{2}\mu^3\p_\phi^2 \big)\tilde\eps+O(\mu^4)\label{eps0}
\eea
and associated with the spin 3 generator
\bea
\Lambda_W & \equiv &(\epsilon,\, \chi)\nn \\
&=&\Big( \mu(\p_\phi^2-{32\mathcal{L}\over3k})-{16\mu^2\over k} \mathcal{W}-\frac{\mu^3}{6}\p_\phi^4 +\frac{32\mu^3}{3k}(\cL \p_\phi^2-3\cL' \p_\phi -\frac{8}{3}\cL'' ),\nn\\
&& 1-{\mu^2\over2}(\p_\phi^2+{32 \mathcal{L}\over3k})  \Big)\tilde \chi+O(\mu^4),
\eea
where $\tilde\eps=\tilde \eps(0,\phi)$, $\tilde\chi=\tilde \chi(0,\phi)$ are the gauge symmmetry parameters on the initial data slice. We will now derive the $\cW_3$ algebra starting from this ansatz.

First, we obtain the conserved charges associated with these symmetry transformations. In order to obtain the conserved charges, we expand the $\p_+$ derivative
as $\p_+ = \frac{1}{2}(\p_t + \p_\phi)$ and we replace all time derivatives acting on the fields and their variations and on the gauge parameters using \eqref{eqLt}-\eqref{eqet}.
The infinitesimal charges are
\bea
\sdelta\cQ_{\Lambda_L} = {1\over2\pi}\int_{t=0} d\phi \,\tilde \epsilon \delta\tilde{\mathcal{L}},\qquad \sdelta\cQ_{\Lambda_W}=-{1\over 2\pi}\int_{t=0} d\phi \, \tilde\chi \delta\tilde{\mathcal{W}}\label{chargeinfi}
\eea
where
\bea
\tilde{\mathcal{L}}&=&\mathcal{L}+3\mu \mathcal{W}+\mu^2 \Big({7\over2}\mathcal{L}''+{16\over3k} \mathcal{L}^2\Big)+\frac{29}{6}\mu^3 \cW'' +O(\mu^4),\\
\tilde{\mathcal{W}}&=&\mathcal{W}-\mu \big(3 \mathcal{L}''-{32\over3k}\mathcal{L}^2\big)+{\mu^2\over2}(3 \mathcal{W}''+\frac{32}{k}\mathcal{L}\mathcal{W})-\frac{512}{27k^2}\mu^3 \cL^3 \nn\\
&& +\frac{16 \mu^3}{9k}\left( 9\cW^2 -66 (\cL')^2-43\cL \cL'' \right)+\frac{77}{18}\mu^3 \cL''''+O(\mu^4).
\eea
Now the infinitesimal charges only depend on the initial data at $t=0$. We are free to choose the gauge symmetry parameters $\tilde\epsilon(0,\phi),\,\tilde\chi(0,\phi)$ independently on the fields $\mathcal{L}(0,\phi),\,\mathcal{W}(0,\phi)$, the charges are then integrable at $t=0$ and we obtain the spin 2 and spin 3 charges
\bea
\cQ^{t=0}_{\Lambda_L} = {1\over2\pi}\int_{t=0} d\phi \,\tilde \epsilon \tilde{\mathcal{L}},\qquad \cQ_{\Lambda_W}^{t=0}=-{1\over 2\pi}\int_{t=0} d\phi \, \tilde \chi \tilde{\mathcal{W}}.
\label{chargefn}\eea
We insist that these charges are only constructed at $t=0$ and allow to obtain the value of the charge associated with any gauge symmetry parameter $\tilde\epsilon(0,\phi),\,\tilde\chi(0,\phi)$ at $t=0$. Since the infinitesimal charges are conserved, the integrability conditions are also conserved, and one can build the charges at another time $t$ by letting time evolve, and integrate the infinitesimal charges at that later time. Given that $(\eps,\chi)$ obey field-dependent evolution laws and the fields themselves obey non-linear partial differential equations, the relationship between $\tilde\epsilon(0,\phi),\,\tilde\chi(0,\phi)$ and $\eps(t,\phi),\chi(t,\phi)$ cannot be easily worked out. We were therefore not able to derive a closed-form expression for the integrated conserved charges associated with $\tilde\epsilon(0,\phi),\,\tilde\chi(0,\phi)$ at $t \neq 0$. Nevertheless, these conserved charges should exist at all times, according to the above reasoning.


Let us now compute the Poisson bracket between the conserved charges using
\bea
\{ \cQ_{\Lambda^{(1)}} ,\cQ_{\Lambda^{(2)}} \} \equiv \delta_{\Lambda^{(1)}} \cQ_{\Lambda^{(2)}} = \sdelta \cQ_{\Lambda^{(2)}}[\delta_{\Lambda^{(1)}} \cL,\delta_{\Lambda^{(1)}} \cW ; \cL(\phi) , \cW(\phi) ]\label{Poi1}
\eea
where ${\Lambda^{(1)}} = (\eps^{(1)},\chi^{(1)})$,  ${\Lambda^{(2)}} = (\eps^{(2)},\chi^{(2)})$ are two choices of generators (either $\Lambda_L$ or $\Lambda_{W}$ with a corresponding choice of $\tilde \eps$ or $\tilde \chi$). Here, the Poisson bracket can be computed using the infinitesimal charge formula given in \eqref{chargephi} even though we do not have at hand the conserved charge $\cQ_\Lambda$ at all times (see \cite{Barnich:2007bf} for a general proof). The infinitesimal charge is linear in the variations of the fields, but it might depend non-linearly on the fields $\cL$ and $\cW$ and their $\phi$ derivatives.  We emphasize that all time dependence has been removed  using \eqref{eqLt}-\eqref{eqet}.

After some algebra, we recognize that the Poisson bracket can equivalently be written as
\bea
\{ \cQ_{\Lambda^{(1)}} ,\cQ_{\Lambda^{(2)}} \} \equiv \frac{1}{2\pi } \int d\phi \left( \tilde\eps^{(1)} \tilde \delta_{\Lambda^{(2)}} \tilde \cL - \tilde\chi^{(1)} \tilde \delta_{\Lambda^{(2)}} \tilde \cW\right)\label{Poi2}
\eea
where
\bea
\tilde \delta_{\Lambda} \tilde \cL &=& -\p_\phi\tilde\cL\tilde\epsilon-2\tilde\cL\p_\phi\tilde\epsilon+{k\over2}\p_\phi^3\tilde \epsilon+2\tilde \chi\p_\phi\tilde\cW+3\p_\phi\tilde\chi\tilde\cW\nn\\
\tilde \delta_{\Lambda} \tilde \cW &=&-\tilde\epsilon\p_\phi\tilde\cW-3\p_\phi\tilde\epsilon\tilde\cW\nn\\
&&-{1\over3}\Big(2\tilde\chi\p_\phi^3\tilde\cL+9\p_\phi\tilde\chi\p_\phi^2\tilde\cL+15\p_\phi^2\tilde\chi\p_\phi\tilde\cL+10\p_\phi^3\tilde\chi\tilde\cL-{k\over2}\p_\phi^5
\tilde\chi \nn\\&&-{32\over k}(\tilde\chi\tilde\cL\p_\phi\tilde\cL+\p_\phi\tilde \chi\tilde\cL^2)\Big)\label{trsftilde}
\eea
is be formally the same  as \eqref{trsf} with $\p_+$ substituted by $\p_\phi$ and $\cL$ by $\tilde \cL$.
However, on the initial data slice, $\tilde\epsilon(0,\phi),\tilde\chi(0,\phi)$ are arbitrary functions of $\phi$ obeying periodic boundary condition in the $\phi$ direction and, moreover, they are independent of the fields $\mathcal{L}(0,\phi),\cW(0,\phi)$. Therefore, we are free to perform a Fourier decomposition, and obtain  the algebra by calculating the Poisson bracket between the Fourier modes.
This Poisson bracket reproduces explicitly the $\cW_3$ algebra. This proves that the canonical charges $\cQ_{\Lambda_L} $ and $\cQ_{\Lambda_W}$ form a $\cW_3$ algebra in perturbation theory in $\mu$.

The same result can be obtained independently in the barred sector with $x^\pm$ exchanged. Since the unbarred and barred sectors mutually commute, the total asymptotic symmetry algebra is therefore $\cW_3 \times \cW_3$.

Let us comment on our results. To understand the meaning of  these symmetry generators in the bulk gravitational theory and its conformal dual, we need a map between the Chern-Simons theory and a metric-like formalism, which can be found in \cite{Campoleoni:2012hp}.
A gauge transformation given by $(\Lambda\equiv \Lambda^AJ_A, \, \bar{\Lambda}\equiv\bar{\Lambda}^AJ_A)$ is associated with a diffeomorphism in the bulk by
\be
\xi^\mu=g^{\mu\nu}\kappa_{AB}e^A_\nu (\Lambda^B-\bar{\Lambda}^B)
\ee
where $J^{A}$ denotes all generators of $sl(3,\mathbb R)$ and $\kappa_{AB}=\frac{1}{2}\; \text{Tr} (J_A J_B)$ is the Killing metric. In general, a Chern-Simons gauge transformation is a combination of local Lorentz-like transformations, diffeomorphisms and spin 3 transformations given by $\Lambda^A-\bar{\Lambda}^A-\xi^\mu e^A_\mu$.
In principle, we can get the asymptotic Killing vectors and accompanying spin-3 transformations associated with all the $\cW_3 \times \cW_3$ generators to any order in $\mu$, $\bar\mu$. Here, we will only look at linear order in both $\mu,\, \bar \mu$. At linear order, one can explicitly show that the Virasoro generator $\tilde{\cL}$ is associated with the Killing vector $\xi=\tilde\epsilon\p_+-{1\over2}\tilde\epsilon'\p_\rho$ which means that it generates a diffeomorphism along $x^+$, combined with a spin 3 transformation. This combination leaves $\bar{\tilde\cL},\,\bar{\tilde \cW}$ invariant. Similarly, $\bar{\tilde{\cL}}$ generates a diffeomorphism along $x^-$ combined with another spin 3 transformation, and the combination of transformations leaves $\tilde\cL,\, \tilde\cW$ invariant. Note that although our generators are only defined at $t=0$, they extend at all times with a specific (field-dependent) dependence on $x^+$ and $x^-$, as the equations of motion tell how they evolve with time. The fact that the Virasoro generators are not associated with pure diffeomorphisms has occured in other situations as well. One example has been discussed in the context of gravity coupled to a $U(1)$ gauge field in $AdS_2$ \cite{Hartman:2008dq}. There, a pure diffeomorphism would not preserve the boundary conditions. Instead, one has improve the stress tensor by doing a (large) $U(1)$ gauge transformation as well. We think that it is exactly what is happening here in the higher spin context: a pure diffeomorphism itself will not preserve the boundary conditions, so it needs to be supplemented with a spin 3 transformation in order to get the correct stress tensor. At quadratic and higher orders in $\mu$, $\bar\mu$, the correct identification of the boundary diffeomorphism would require more care since the leading asymptotic behavior of the metric changes.

\subsection{Virasoro zero modes}
\label{zeromodes}

Let us discuss in more detail the zero modes of the Virasoro algebra.

First, one can build from the infinitesimal charges \eqref{chargephi} the integrable charges associated with $(\eps,\chi)= (1,0)$ and $(\eps,\chi) = (0,1)$ since these gauge parameters are solutions to the system \eqref{epchi}. We obtain
\bea
\cQ_{(1,0)} &=& \frac{1}{2\pi} \int d\phi \left( \cL +2\mu \cW \right),\nn\\
\cQ_{(0,1)} &=& \frac{1}{2\pi} \int d\phi  \left(-\cW + \frac{2\mu}{3} ( \p_+^2 \cL - \frac{8}{k}\cL^2) \right).\label{chargesy}
\eea
A similar expression holds for $(\bar \eps,\bar \chi) = (1,0)$ and $(0,1)$ with all quantities barred and $\p_+$ derivatives exchanged with $\p_-$ derivatives. These charges are conserved in time and commute under the Poisson bracket.

The charges associated with the asymptotic Killing vectors $\p_\pm$ can also be obtained from \eqref{chargephi} upon setting $\Lambda = \xi^\mu A_\mu$, $\bar \Lambda = \xi^\mu \bar A_\mu$ (see discussions in \cite{Campoleoni:2012hp}). The asymptotic Killing vector $\p_+$ corresponds to $(\eps,\chi,\bar\eps,\bar\chi)=(1,0,0,\bar \mu)$ and $\p_-$ corresponds to $(\eps,\chi,\bar\eps,\bar\chi)=(0,\mu,1,0)$. Evaluating on the fixed $t$ slice, the charges are
\bea \Delta&\equiv&\cQ_{\p_+}= {1\over2\pi}\int_0^{2\pi} d\phi \Big( \mathcal{L}+2\mu \mathcal{W}-\bar{\mu} \bar{\mathcal{W}}-{16\over 3k}\bar{\mu}^2\bar {\mathcal{L}}^2+{\bar{ \mu}^2\over 6}\p_t^2\bar{\mathcal{L}}\Big),\nn\\
\bar{\Delta}&\equiv&\cQ_{\p_-}={1\over2\pi}\int_0^{2\pi}  d\phi \Big( \bar{\mathcal{ L}}+2\bar{\mu} \bar{\mathcal{W}}-\mu \mathcal{W}-{16\over 3k}{\mu}^2 {\mathcal{L}}^2+{{ \mu}^2\over 6}\p_t^2{\mathcal{L}}\Big), \label{energyphi}
\eea
which agree with the computation of \cite{Perez:2012cf} for time-independent solutions in the phase space where $\mu$ is fixed.

Let us now compute the zero modes of the Virasoro algebra. From \eqref{eps0}, the unbarred zero mode Virasoro generator is associated with $(\eps,\chi,\bar\eps,\bar\chi)=(1,-\mu,0,0)$ while the barred zero mode Virasoro generator is associated with $(\eps,\chi,\bar\eps,\bar\chi) = (0,0,1,-\bar\mu)$. They are given by
\bea
\tilde \Delta&\equiv&\cQ_{(1,-\mu,0,0)}= {1\over2\pi}\int_0^{2\pi} d\phi \Big( \mathcal{L}+3\mu \mathcal{W} + {16\over 3k}\mu^2\cL^2 +O(\mu^4) \Big),\\
\bar{\tilde{\Delta}}&\equiv&\cQ_{(0,0,1,-\bar\mu)}= {1\over2\pi}\int_0^{2\pi} d\phi \Big(\bar\cL+3\bar\mu \bar\cW + {16\over 3k}\bar\mu^2\bar\cL^2+O(\mu^4) \Big), \label{Virtilde0}
\eea
at the initial time $t=0$. These expressions can be obtained either from the definition \eqref{chargefn} for $\tilde\eps =1$ or from expanding the appropriate linear combination of \eqref{chargesy} using the equations of motion \eqref{eqLt}.

Let us comment on our result. The Virasoro zero modes $\tilde \Delta$, $\bar{\tilde{\Delta}}$ do not agree with the naive left and right-moving generators $\Delta$, $\bar \Delta$. The reason is that upon turning on $\mu$ and $\bar \mu$, the unbarred Virasoro generator starts to be also slightly left-moving and the barred Virasoro generator starts to be slightly right-moving. There is however no mixing between the unbarred and barred sectors since the $sl(3,\mathbb R)\times sl(3,\mathbb R)$ Chern-Simons theory (including the boundary terms) is simply the sum of the two uncoupled unbarred and barred theories. Interestingly, the difference between the unbarred and barred Virasoro zero modes
\bea
\tilde \Delta - \bar{\tilde{\Delta}} = \Delta - \bar{\Delta} +O(\mu^4) \equiv J + O(\mu^4)
\eea
agrees with the angular momentum $J$, at least up to $O(\mu^4)$. One argument for such a conservation under $\mu,\bar\mu$ deformation is that in the semi-classical theory the angular momentum is quantized and cannot therefore be changed with a continuous parameter. On the contrary, nothing prevents the expression for the energy to change and its expression is indeed affected upon turning on $\mu,\bar \mu$.  It would be of course interesting to reproduce the expression for the Virasoro zero modes from the dual holographic theory.


\subsection{KdV charges}
\label{sec:can}

Let us now take another perspective on the conserved charges analysis. By inspection, we can identify at least four linearly independent solutions to the gauge parameter equations \eqref{epchi} around a generic point in phase space. Two solutions are the obvious constant parameters  $(\eps,\chi)= (1,0)$ and $(\eps,\chi) = (0,1)$ whose charges have been obtained in the last section. Two non-trivial solutions are given by
\bea
\left( \begin{array}{c} \eps \\ \chi \end{array} \right)= \left( \begin{array}{c} \cW \\ -\cL \end{array} \right), \; \left( \begin{array}{c} \frac{1}{3} (\p_+^2 \cL - \frac{8}{k}\cL^2) \\ \cW \end{array} \right).\label{four}
\eea
For phase space elements with constant $\cL$ and $\cW$, some of these symmetries degenerate but they are independent in general.  The two charges associated with \eqref{four} are integrable and given by
\bea
\cQ_{(\cW,-\cL)} &=& \frac{1}{2\pi} \int d\phi \left( \cW \cL + \frac{2\mu}{3} (\frac{3}{2}\cW^2+\frac{1}{2}(\p_+ \cL)^2\right.\nn\\
&&\left.  - \cL \p_+^2 \cL + \frac{16}{3k}\cL^3)  \right),\nn\\
 \cQ_{(\frac{1}{3} (\p_+^2 \cL - \frac{8}{k}\cL^2),\cW)} &=& \frac{1}{2\pi} \int d\phi  \left(- \frac{\cW^2}{2}-\frac{8}{9k}\cL^3 - \frac{1}{6}(\p_+ \cL)^2 \nn \right. \\
&&\left. + \frac{2 \mu}{3} ( \cW (\p_+^2 \cL - \frac{8}{k} \cL^2 )   -\p_+ \cW \p_+ \cL) \right) .\label{nonlc}
\eea
In order to obtain the last expression we used the equations of motion and we performed an integration by parts in $\p_\phi$. We can check that they are conserved on-shell at all times.

Let us make contact with the KdV integrable hierarchy of charges \eqref{Ck}. These charges are defined using $t^+$ as time evolution parameter. We can however reformulate the conservation laws \eqref{cons} after using our variables $x^\pm = t \pm \phi$ as
\bea
\p_t (h_k-2\mu l_k) = \p_\phi (h_k+2\mu\, l_k).
\eea
Therefore, we can build the hierarchy of conserved quantities under $t$ evolution from
\bea
\cH_k = \int d\phi (h_k-2\mu l_k),\qquad k \in \mathbb N, \quad k\neq 3\mathbb N.\label{Hk}
\eea
Since $h_k$ and $l_k$ are $\mu$ independent, the conserved charges are linear in $\mu$. Using \eqref{hk}, the first four charges in the $n=3$ KdV hierarchy exactly reproduce the four conserved charges that we derived using canonical methods in \eqref{chargesy}-\eqref{nonlc}. We expect that one can reproduce the entire integrable tower of conserved charges from suitable field-dependent solutions to \eqref{epchi}, but it remains to be proven.

In the standard KdV hierarchy, all charges \eqref{Ck} commute, which provides precisely with the integrability structure. Here, the Poisson bracket of the conserved charges \eqref{chargesy}-\eqref{nonlc} can be computed using the canonical bracket
\bea
\{ \cQ_{(\eps,\chi)} ,  \cQ_{(\eps',\chi')} \} = \delta_{(\eps,\chi)}\cQ_{(\eps',\chi')}
\eea
with the variation of the fields \eqref{trsf}. After an involved but straightforward computation, the result is that the four charges commute for any $\mu$.

 It would be interesting to investigate if all charges \eqref{Hk} commute, maybe using the definition of $l_k$ \eqref{lk} in terms of Lax operators. We leave this issue for future investigations. It has been proposed that the tower of commuting KdV charges \eqref{Ck} can be viewed as the Cartan subalgebra of a linear extensions of the $\cW_3$ algebra, $\cW_3^{lin}$ \cite{Malik:1996rb}. It would be interesting to interpret the charges \eqref{Hk} in that framework as well.

\section{$\cW^{(2)}_3$ symmetry in the Diagonal embedding}

In this section we discuss the diagonal embedding. We will derive the canonical charges at finite $\lambda$ and show that they obey the $\cW_3^{(2)}$ algebra under the canonical Poisson bracket. We will then obtain the bi-Hamiltonian structure of the Boussinesq equation in $t$ evolution which describes the dynamics of the phase space in the diagonal embedding. We will show that the second Hamiltonian structure  precisely coincides with the $\cW_3^{(2)}$ Hamiltonian structure after a field redefinition.

\subsection{Canonical analysis}
In this subsection we perform a canonical analysis for the diagonal embedding. As mentioned before, there are four initial data $\mathcal{J},\mathcal{G}_\pm,\mathcal{T}$ at $t=0$.
Under the field redefinition \eqref{fr}, these variables are related to $\mathcal{L},\mathcal{W},\dot{\mathcal{L}},\ddot{\mathcal{L}}$ at $t=0.$
Similarly,
the gauge transformations are determined by the initial data of the four
 new variables $\varepsilon,\,\eta,\,\alpha_\pm$ (that we choose in order to obtain normalized charges at $\lambda =0$ as we will see shortly) as
\bea
\chi&=&{1\over 4\lambda^2} \varepsilon,\\
\epsilon&=&{1\over2\sqrt{2}\lambda}(\alpha_++\alpha_-),\\
\p_+\chi&=&-{1\over2\sqrt{2}\lambda}(\alpha_+-\alpha_-),\\
\p_+^2\chi&=&-\eta.
\eea
We warn the reader that we introduce the new symbol $\varepsilon$ different than $\eps$. It follows from the equations of motion for $\epsilon$ and $\chi$ (\ref{epchi}) that
the new variables satisfy the following linear differential equations
\bea
\dot{\varepsilon}&=&-\varepsilon'-2\sqrt{2}\lambda (\alpha_+-\alpha_-),\nn\\
\dot{\alpha}_+&=&-\alpha'_+-{2\lambda}\Big(\lambda(\alpha_+-\alpha_-)+{1\over \sqrt{2}}(\varepsilon'-2\eta)\Big),\nn\\
\dot{\alpha}_-&=&-\alpha'_--{2\lambda}\Big(\lambda(\alpha_+-\alpha_-)+{1\over \sqrt{2}}(\varepsilon'+2\eta)\Big),\\
\dot{\eta}&=&-\eta'-{3\sqrt{2}}\lambda\Big(2(8\frac{\mathcal{J}}{k}+\lambda^2 )(\alpha_+-\alpha_-)+\sqrt{2}\lambda \varepsilon'+(\alpha'_++\alpha'_-)\Big),\nn
\eea
where $\cJ$ is related to $\cL$ by \eqref{frtwo}.
The transformation rules of the $\mathcal{L}$ and $\mathcal{W}$ variables can be also translated into transformation rules of $\mathcal{J},\mathcal{G}_\pm,\mathcal{J}$.
The infinitesimal conserved charge \eqref{chargephi} associated with the gauge parameters $K\equiv(\eta,\,\alpha_+,\,\alpha_-,\,\varepsilon)$ can then be written as
\bea
\sdelta \cQ_K &=& {1\over 2\pi}\int d\phi \Big(\varepsilon(\delta\mathcal{T}-{\lambda\over\sqrt{2}}\delta (\mathcal{G}_++\mathcal{G}_-)) +\eta\delta\mathcal{J}+\alpha_+(\delta \mathcal{G}_+-{3\over\sqrt{2}}\lambda \delta\mathcal{J})\nn\\
&& +\alpha_-(\delta \mathcal{G}_--{3\over\sqrt{2}}\lambda \delta\mathcal{J}) \Big).
\eea
It is linear and it can then be directly integrated at $t=0$ yielding
\bea
\cQ_K &=& {1\over 2\pi}\int_{t=0} d\phi \Big(\varepsilon(\mathcal{T}-{\lambda\over\sqrt{2}} (\mathcal{G}_++\mathcal{G}_-)) +\eta\mathcal{J}+\alpha_+( \mathcal{G}_+-{3\over\sqrt{2}}\lambda\mathcal{J})\nn\\
&& +\alpha_-(\mathcal{G}_--{3\over\sqrt{2}}\lambda\mathcal{J}) \Big).
\eea
One can check that the following charges
\bea \cQ_\eta&=& {1\over 2\pi}\int_{t=0} d\phi \ \eta(\mathcal{J}+{k\over2}\lambda^2)\equiv {1\over 2\pi}\int d\phi \,\eta\mathcal{J}_\lambda\label{w32charge}\\
\cQ_{\alpha_\pm}&=&{1\over 2\pi}\int_{t=0} d\phi  \ \alpha_\pm \Big(\mathcal{G}_\pm-\lambda(3\sqrt{2}\mathcal{J}+{k\over \sqrt{2}}\lambda^2)\Big)\equiv{1\over2\pi} \int d\phi \,\alpha_\pm \mathcal{G}_{\pm \lambda}\nn\\
\cQ_\varepsilon&=&{1\over 2\pi}\int_{t=0} d\phi  \ \varepsilon \big(\mathcal{T}-{3\lambda\over\sqrt{2}}(\mathcal{G}_++\mathcal{G}_-)+6\lambda^2 \mathcal{J}\big)
\equiv{1\over 2\pi}\int_{t=0} d\phi\  \varepsilon\mathcal{T}_\lambda\nn
\eea
associated with the gauge parameters
\bea
K_\eta&=&(\eta,0,0,0),\\
K_{\alpha_+}&=&(-{3\over\sqrt{2}}\lambda\alpha_+,\alpha_+,0,0),\\
K_{\alpha_-}&=&(-{3\over\sqrt{2}}\lambda\alpha_-,0,\alpha_-,0),\\
K_\varepsilon&=&(0,-\sqrt{2}\lambda\varepsilon,-\sqrt{2}\lambda\varepsilon,\varepsilon),
\eea
form the $\cW^{(2)}_3$ algebra. The charges $\cQ_\varepsilon$ form a Virasoro algebra with central charge $\hat{c}={3k\over2}$, which is unchanged from $\lambda =0$ \cite{Ammon:2011nk}. The charges $\cQ_\eta$ form a U(1) algebra with level $\hat{k}=-{k\over3}$, which is unchanged from $\lambda = 0$ \cite{Ammon:2011nk,Castro:2012bc}.

\subsection{Bi-Hamiltonian structure}
\label{biH2}

We now construct the bi-Hamiltonian structure of the Boussinesq system \eqref{sol2} at constant $t$, with $\phi \sim \phi +2\pi$. We will largely follow the work of Mathieu and Oevel \cite{Mathieu:1991et}.

The crucial ingredient which allows to build the Hamiltonian structure is the Miura map, which provides with a free field realization of non-linear algebras. The Miura transformation of the $Bsq$ equation can be written as
\bea
u &=& p_{1x}-\frac{1}{2}(p_1^2+p_2^2),\nn \\
v &=& s - p_{2xx}+3 p_1 p_{2x}+p_{1x}p_2+\frac{2}{3}p_2^3-2p_1^2 p_2\label{Miura}
\eea
where $s$ is an arbitrary constant parameter and $x \equiv x^+$. This transformation maps solutions of the Boussinesq equation to the solutions of the modified Boussinesq (mBsq) equation
\bea
\p_{\hat x^-} \left( \begin{array}{c} p_1 \\ p_2 \end{array} \right) =  \left( \begin{array}{c} -p_{2xx}+2(p_1 p_2)_x \\ p_{1xx}+(p_1^2-p_2^2)_x \end{array} \right) .
\eea

We can write the mBsq equation in terms of our time and angle $x^\pm = t \pm \phi$ after defining
 \bea
 q_1=\p_+ p_1,\quad q_2=\p_+p_2\, ,\label{defq}
 \eea
as
\bea
\p_t \left(
\begin{array}{ccc}
  p_1  \\
  p_2 \\
 q_1\\
 q_2
\end{array}
\right)
=\left(
\begin{array}{ccc}
2q_1-p_1'    \\
 2q_2 -p_2' \\
 {\sqrt{3}\over \mu}(q_2-p_2')-4p_1q_1+4p_2q_2-q_1'\\
 -  {\sqrt{3}\over \mu}(q_1-p_1')+4p_2q_1+4p_1q_2-q_2'
\end{array}
\right)
\eea
where primes denote $\p_\phi$ derivatives. In order to simplify the system, we introduce the new fields $\phi_1$, $\phi_2$ from the field redefinition
\bea
\phi_1=q_1+p_1^2-p_2^2-{\sqrt{3}\over2\mu}p_2,\quad \phi_2=q_2-2p_1p_2+{\sqrt{3}\over2\mu}p_1.\label{defphi}
\eea
Then
\bea\p_t
\left(
\begin{array}{ccc}
  p_1  \\
  p_2 \\
 \phi_1\\
 \phi_2
\end{array}
\right)
=\left(
\begin{array}{ccc}
2(\phi_1-p_1^2+p_2^2+{\sqrt{3}\over2\mu}p_2)-p_1'\\ 2(\phi_2+2p_1p_2-{\sqrt{3}\over2\mu}p_1)-p_2' \\ -{\sqrt{3}\over\mu}p_2'-\phi_1' \\
{\sqrt{3}\over\mu}p_1'-\phi_2'
\end{array}
\right).\label{EOMpphi}
\eea
We then observe by inspection that the system can be written as a Hamiltonian system,
\bea
\frac{k\mu }{\sqrt{3}} \, \p_{t} \left( \begin{array}{c} p_1 \\ p_2 \\ \phi_1 \\ \phi_2 \end{array} \right) = {\Theta} \left( \begin{array}{c} \frac{\delta }{\delta p_1}\\   \frac{\delta }{\delta p_2}\\  \frac{\delta }{\delta \phi_1}\\  \frac{\delta }{\delta \phi_2} \end{array} \right)
 \mathcal H ,
 \eea
 with
\bea
{\Theta} &=& \left(
\begin{array}{cccc}
 0&-1&0&0  \\
  1&0&0&0 \\
 0&0&{\sqrt{3}\over2\mu}\p_\phi&0\\
 0&0&0&{\sqrt{3}\over2\mu}\p_\phi
\end{array}
\right),\\
\mathcal H &=& \frac{k\mu }{\sqrt{3}}  \int_0^{2\pi} d\phi \left(2(p_1\phi_2-p_2\phi_1+p_1^2p_2-{1\over3}p_2^3)-p_1p_2' \right. \nn\\
&&\left. -{\sqrt{3}\over2\mu}(p_1^2+p_2^2)-{\mu\over\sqrt{3}}(\phi_1^2+\phi_2^2)\right) .\label{HamBH}
\eea

We recognize the second Hamiltonian as the  ``unbarred connection part'' $\cQ_{(1,\mu)} $ of the canonical energy $\cQ_{\p_t}$ derived in Section \ref{zeromodes},
\bea
 \cH +\frac{s}{2} &=&\cQ_{(1,\mu)} \nn\\
&=& \frac{1}{2\pi } \int_0^{2\pi} d\phi \left( \cL +\mu \cW + \frac{2\mu^2}{3} (\p_+^2 \cL- \frac{8}{k}\cL^2)\right)\nn \\
&=&\frac{1}{2\pi} \int_0^{2\pi} d\phi  \left( \cT +\frac{\lambda}{\sqrt{2}}(\cG^+ +\cG^-) -6\lambda^2 \cJ \right)
\eea
up to a trivial shift. Since the canonical Poisson bracket corresponds to the second Hamiltonian structure, we need to set $s = 0$ in order to compare the formalisms.
It is natural that the canonical energy is precisely the second Hamiltonian of the integrable hierarchy which generates the Poisson bracket.

We can now formulate the Hamiltonian structure for the original Boussinesq equation written as an evolution along $t$. Starting from \eqref{sol2} and splitting $t$ and $\phi$, we can write the dynamics in terms of $(\cJ,\cG^\pm,\cT)$ as
\bea
\p_{t} \left( \begin{array}{c} \cJ \\ \cG^+ \\\cG^- \\ \cT  \end{array} \right) =  \left( \begin{array}{c}-2\sqrt{2}\lambda (\cG^+-\cG^-) -\p_\phi \cJ \\ -6\lambda^2 (\cG^+ - \cG^-)+2\sqrt{2} \lambda \cT -3 \sqrt{2}\lambda \p_\phi \cJ - \p_\phi \cG^+ + \frac{24\sqrt{2}}{k}\lambda \cJ^2 \\
 -6\lambda^2 (\cG^+ - \cG^-)-2\sqrt{2} \lambda \cT -3 \sqrt{2}\lambda \p_\phi \cJ - \p_\phi \cG^- - \frac{24\sqrt{2}}{k}\lambda \cJ^2\\
 -6\sqrt{2}\lambda^3 (\cG^+ - \cG^-)-\sqrt{2} \lambda \p_\phi (\cG^+ + \cG^-)-6\lambda^2 \p_\phi \cJ - \p_\phi \cT
 \end{array} \right).\nn
\eea
The Miura map \eqref{Miura}, the definitions \eqref{defuv}-\eqref{defq}-\eqref{defphi} and the field equations \eqref{EOMpphi} allow to express the fields in terms of $p_1,p_2,\phi_1,\phi_2$ as
\bea
u \equiv \frac{12\lambda^2}{k}\cJ &=& 2\sqrt{3} \lambda^2 p_2 -\frac{1}{2}(3p_1^2-p_2^2-2\phi_1),\\
v \equiv \frac{4\sqrt{6}\lambda^3}{k}(\cG^+ + \cG^- )&=&12 \lambda^4p_2 + 2\sqrt{3}\lambda^2 (\phi_1-2p_1^2-\p_\phi p_1)\nn\\
&& \hspace{-1cm}+(s+p_1 \phi_2 - p_2 \phi_1 +p_1^2 p_2-\frac{1}{3}p_2^3),\\
 w \equiv -\frac{12 \sqrt{2}\lambda^3}{k}(\cG^+ - \cG^-)&=& -12\lambda^4 p_1 + 2\sqrt{3}\lambda^2 (\phi_2-2p_1 p_2 -\p_\phi p_2)\nn\\
&&\hspace{-1cm}+ (-3p_1\phi_1 +p_2 \phi_2 -p_1 p_2^2 +3p_1^3),\\
z+12\lambda^4 u\equiv -\frac{48\lambda^4}{k}(\cT-3\lambda^2\mathcal{J}) &=&18\lambda^4 (p_1^2+p_2^2)+24\lambda^4 \p_\phi p_1 + 2\sqrt{3}\lambda^2(  \nn\\
&&\hspace{-6.5cm} -3p_1^2 p_2+p_2^3+3p_2\phi_1 -3p_1 \phi_2 -p_2 \p_\phi p_1+p_1 \p_\phi p_2 -\p_\phi \phi_2)+(\phi_1^2+\phi_2^2).
\eea
The Fr\'echet derivative of $ U \equiv ( u, v, w, z)$ with respect to $P \equiv (p_1,p_2,\phi_1,\phi_2)$ is a $4 \times 4$ matrix operator whose components are given by
\bea
{\bf D}_{ij} &=& \frac{\p}{\p P_j} U_i + \left( \frac{\p}{\p \p_\phi P_j}  U_i\right) \p_\phi \equiv {\bf D}^{[0]}_{ij} +  {\bf D}^{[1]}_{ij}\p_\phi\,
\eea
where $i,j=1,2,3,4$. Its adjoint is given by
\bea
{\bf D}^+ =\left(  {\bf D}^{[0]} -\p_\phi  {\bf D}^{[1]} - {\bf D}^{[1]}\p_\phi \right)^T
\eea
where $T$ is the transpose.
The two Hamitonian operators $\Theta^1$, ${\Theta}^2$ are then obtained as
\bea
{ \Theta}^2 + s\,  {\Theta}^1 = {\bf D} \; {\Theta}\;  {\bf D}^+
\eea
The result for  $ \Theta^1$, ${\Theta}^2$, that we don't find particularly useful to display here, is considerably simplified after performing the field redefinition
\bea
\tilde u &=& \frac{1}{4\sqrt{3}\mu} (3+8\mu^2u),\\
\tilde v &=& -\frac{1}{2\sqrt{2}3^{1/4}\mu^{3/2}}(1+8\mu^2 u - {8\over\sqrt{3}}\mu^3  v),\\
\tilde w &=& \frac{2\sqrt{2}\mu^{3\over2}}{3^{3/4}} w,\\
\tilde z &=& -2u+2\sqrt{3}\mu v+{4\mu^2\over3}z-2\sqrt{3}\mu s.
\eea
We denote the Jacobian of the transformation of the tilde variables $\tilde u,\tilde v,\tilde w,\tilde z $ in terms of the variables $u, v, w, z$ as ${\bf J}$. We then recognize the first and second Hamiltonian structure
\bea
{\tilde \Theta}^1 &=&   {\bf J} \, { \Theta}^1 \, {\bf J}^T \nn\\
&=& -\frac{4\sqrt{2}}{3^{1/4}}\mu^{5/2} \left(
\begin{array}{cccc}
0&0&1&0\\
0&0&0&-\p_\phi\\
-1&0&0&0\\
0&-\p_\phi &0&0
 \end{array}
\right) , \\
{\tilde \Theta}^2 &=&  {\bf J} \, { \Theta}^2 \, {\bf J}^T \nn\\
&=&-\frac{2\mu}{\sqrt{3}} \left(
\begin{array}{cccc}
-\p_\phi & \tilde w & 3 \tilde v & -2 \tilde u_\phi-2\tilde u \p_\phi \\
-\tilde w & \tilde u_\phi+2\tilde u \p_\phi & \p_\phi^2 -\tilde z+4\tilde u^2 & -2 \tilde v_\phi -3\tilde v\p_\phi \\
-3 \tilde v &-\p_\phi^2 + \tilde z - 4\tilde u^2 & -3\tilde u_\phi - 6 \tilde u\p_\phi & -2\tilde w_\phi - 3\tilde w\p_\phi \\
-2\tilde u \p_\phi & -\tilde v_\phi -3 \tilde v \p_\phi & - \tilde w_\phi - 3 \tilde w \p_\phi &\p_\phi^3 -2\tilde z_\phi - 4\tilde z \p_\phi
 \end{array}
\right),\nn
\eea
as the one of the $\cW_3^{(2)}$ algebra \eqref{th12}-\eqref{th22} up to an irrelevant multiplicative constant.

The standard spin 2, 3/2 and 1 generators of the $\cW_3^{(2)}$ algebra $(\cJ_\lambda,\cG_\lambda^\pm, \cT_\lambda)$ at any finite value of $\lambda = 1/(2\mu^{1/2})$ are proportional, respectively,
to $\tilde u$, $\tilde v \pm \frac{1}{\sqrt{3}} \tilde w$ and $\tilde z$. We can find the prefactors by matching the result at $\lambda = 0$ and we find
\bea
\cJ_\lambda &\equiv & \frac{k}{2\sqrt{3}}\tilde u  = \cJ+\frac{k}{2}\lambda^2 ,\\
\cG_\lambda^\pm & \equiv & -\frac{3^{1/4}}{4}k(\tilde v \pm \frac{1}{\sqrt{3}} \tilde w)=  \cG^\pm-3\sqrt{2}\lambda \cJ-\frac{k}{\sqrt{2}}\lambda^3 ,\\
\cT_\lambda & \equiv & -\frac{k}{4}\tilde z-\frac{\sqrt{3}k}{8\lambda^2}s =  \cT -\frac{3}{\sqrt{2}}\lambda (\cG^+ + \cG^-)+6\lambda^2 \cJ.
\eea
We see that these generators exactly agree with the generators we found via the canonical analysis (\ref{w32charge}) upon setting the shift $s=0$. In conclusion, both the canonical and integrability formalisms agree and lead to the $\cW_3^{(2)}$ symmetry algebra at any finite $\lambda$. The preservation of the $\cW_3^{(2)}$ integrability structure under deformation was also noticed in \cite{DeGroot:1991ca, Burroughs:1991bd}.

\section*{Acknowledgements}
We are grateful to A.~Castro,  C.~Chang, S.~Fredenhagen, M.~Gaberdiel, M.~Henneaux, J.~Jottar, P.~Mathieu, E. Perlmutter and C. Troessaert for stimulating discussions and useful comments. We also thank I. Krichever for pointing out relevant references. G.C. also thank J. Jottar for sharing his notes on Chern-Simons theory. This work is partly supported by NSF grant 1205550. W.S. is supported in part by the Harvard Society of Fellows, G.C. is a Research Associate of the Fonds de la Recherche Scientifique F.R.S.-FNRS (Belgium).
W.S. thanks the Galileo Galilei Institute for Theoretical Physics for their hospitality and the INFN for partial support. G.C. also thanks the Michell Institute for Theoretical Physics at Texas A\&M for their kind hospitality.

\appendix

\section{Conventions}
\label{conv}

The $sl(3,\mathbb R)$ generators in the principal $sl(2,\mathbb R)$ embedding are denoted as $L_{\pm 1},\, L_0$, $W_{\pm 2}$, $W_{\pm 1}$, $W_0$. They obey the following commutation relations
\bea
\,[L_i,L_j] &=&(i-j)L_{i+j},\nn\\
\,[L_i,W_m]&=&(2i-m)W_{i+m},\\
\,[W_m,W_n]&=&-\frac{1}{3}(m-n)(2m^2+2n^2-m n-8)L_{m+n}\nn
\eea
where $i,j=-1,0,1$, $m,n=-2,-1,0,1,2$. The $sl(3,\mathbb R) $ algebra in the diagonal embedding is generated by $\hat L_{\pm 1},\hat L_0$, $\hat G^\pm_{\pm 1/2}$ and $\hat J_0$. The two sets of $sl(3,\mathbb R)$ generators expressed in a form convenient for each embedding are simply related by the field redefinition
\bea
\hat L_0 = \frac{1}{2}L_0,\qquad \hat L_{\pm 1} = \pm \frac{1}{4} W_{\pm 2}, \qquad \hat J_0 = \frac{1}{2}W_0,\\
\hat G_{1/2}^\pm = \frac{1}{\sqrt{8}} (W_1 \mp L_1),\qquad \hat G^\pm_{-1/2} = \frac{1}{\sqrt{8}}(L_{-1} \pm W_{-1}).\label{fieldr}
\eea
These conventions follow from the ones of the appendix of \cite{Kraus:2011ds} but with $q=1/2$. We use the trace relations
\bea
&&\text{Tr}(L_{-1} L_1)=-1,\qquad \text{Tr}(L_0 L_0 ) = \frac{1}{2},\nn \\
&&\text{Tr}(W_0 W_0) = \frac{2}{3},\qquad \text{Tr}(W_1 W_{-1}) = -1,\qquad \text{Tr}(W_2 W_{-2}) = 4.
\eea
All other traces vanish.


\begin{thebibliography}{10}

\bibitem{Vasiliev:1995dn}
M.~A. Vasiliev, ``{Higher spin gauge theories in four-dimensions,
  three-dimensions, and two-dimensions},'' {\em Int.J.Mod.Phys.} {\bf D5}
  (1996) 763--797,
\href{http://www.arXiv.org/abs/hep-th/9611024}{{\tt hep-th/9611024}}.

\bibitem{Vasiliev:1996hn}
M.~A. Vasiliev, ``{Higher spin matter interactions in (2+1)-dimensions},''
\href{http://www.arXiv.org/abs/hep-th/9607135}{{\tt hep-th/9607135}}.

\bibitem{Blencowe:1988gj}
M.~Blencowe, ``A consistent interacting massless higher spin field theory in d
  = (2+1),'' {\em Class.Quant.Grav.} {\bf 6} (1989)
443.

\bibitem{Bergshoeff:1989ns}
E.~Bergshoeff, M.~Blencowe, and K.~Stelle, ``Area preserving diffeomorphisms
  and higher spin algebras,'' {\em Commun.Math.Phys.} {\bf 128} (1990)
213.

\bibitem{Gaberdiel:2010pz}
M.~R. Gaberdiel and R.~Gopakumar, ``{An $AdS_3$ Dual for Minimal Model CFTs},''
  {\em Phys.Rev.} {\bf D83} (2011) 066007,
\href{http://www.arXiv.org/abs/1011.2986}{{\tt 1011.2986}}.

\bibitem{Ahn:2011pv}
C.~Ahn, ``{The Large N 't Hooft Limit of Coset Minimal Models},'' {\em JHEP}
  {\bf 1110} (2011) 125,
\href{http://www.arXiv.org/abs/1106.0351}{{\tt 1106.0351}}.

\bibitem{Chang:2011mz}
C.-M. Chang and X.~Yin, ``{Higher Spin Gravity with Matter in AdS$_3$ and Its
  CFT Dual},'' {\em JHEP} {\bf 1210} (2012) 024,
\href{http://www.arXiv.org/abs/1106.2580}{{\tt 1106.2580}}.

\bibitem{Gaberdiel:2011nt}
M.~R. Gaberdiel and C.~Vollenweider, ``{Minimal Model Holography for SO(2N)},''
  {\em JHEP} {\bf 1108} (2011) 104,
\href{http://www.arXiv.org/abs/1106.2634}{{\tt 1106.2634}}.

\bibitem{Creutzig:2011fe}
T.~Creutzig, Y.~Hikida, and P.~B. Ronne, ``{Higher spin AdS$_3$ supergravity
  and its dual CFT},'' {\em JHEP} {\bf 1202} (2012) 109,
\href{http://www.arXiv.org/abs/1111.2139}{{\tt 1111.2139}}.

\bibitem{Gaberdiel:2012ku}
M.~R. Gaberdiel and R.~Gopakumar, ``{Triality in Minimal Model Holography},''
  {\em JHEP} {\bf 1207} (2012) 127,
\href{http://www.arXiv.org/abs/1205.2472}{{\tt 1205.2472}}.

\bibitem{Creutzig:2012ar}
T.~Creutzig, Y.~Hikida, and P.~B. R¿nne, ``{N=1 supersymmetric higher spin
  holography on AdS$_3$},'' {\em JHEP} {\bf 1302} (2013) 019,
\href{http://www.arXiv.org/abs/1209.5404}{{\tt 1209.5404}}.

\bibitem{Chang:2013izp}
C.-M. Chang and X.~Yin, ``{A semi-local holographic minimal model},''
\href{http://www.arXiv.org/abs/1302.4420}{{\tt 1302.4420}}.

\bibitem{Beccaria:2013wqa}
M.~Beccaria, C.~Candu, M.~R. Gaberdiel, and M.~Groher, ``{N=1 extension of
  minimal model holography},''
\href{http://www.arXiv.org/abs/1305.1048}{{\tt 1305.1048}}.

\bibitem{Gaberdiel:2013vva}
M.~R. Gaberdiel and R.~Gopakumar, ``{Large N=4 Holography},''
\href{http://www.arXiv.org/abs/1305.4181}{{\tt 1305.4181}}.

\bibitem{Kraus:2011ds}
P.~Kraus and E.~Perlmutter, ``{Partition functions of higher spin black holes
  and their CFT duals},'' {\em JHEP} {\bf 1111} (2011) 061,
\href{http://www.arXiv.org/abs/1108.2567}{{\tt 1108.2567}}.

\bibitem{Gaberdiel:2011zw}
M.~R. Gaberdiel, R.~Gopakumar, T.~Hartman, and S.~Raju, ``{Partition Functions
  of Holographic Minimal Models},'' {\em JHEP} {\bf 1108} (2011) 077,
\href{http://www.arXiv.org/abs/1106.1897}{{\tt 1106.1897}}.

\bibitem{Ammon:2011ua}
M.~Ammon, P.~Kraus, and E.~Perlmutter, ``{Scalar fields and three-point
  functions in D=3 higher spin gravity},'' {\em JHEP} {\bf 1207} (2012) 113,
\href{http://www.arXiv.org/abs/1111.3926}{{\tt 1111.3926}}.

\bibitem{Papadodimas:2011pf}
K.~Papadodimas and S.~Raju, ``{Correlation Functions in Holographic Minimal
  Models},'' {\em Nucl.Phys.} {\bf B856} (2012) 607--646,
\href{http://www.arXiv.org/abs/1108.3077}{{\tt 1108.3077}}.

\bibitem{Chang:2011vka}
C.-M. Chang and X.~Yin, ``{Correlators in $W_N$ Minimal Model Revisited},''
  {\em JHEP} {\bf 1210} (2012) 050,
\href{http://www.arXiv.org/abs/1112.5459}{{\tt 1112.5459}}.

\bibitem{Castro:2011zq}
A.~Castro, M.~R. Gaberdiel, T.~Hartman, A.~Maloney, and R.~Volpato, ``{The
  Gravity Dual of the Ising Model},'' {\em Phys.Rev.} {\bf D85} (2012) 024032,
\href{http://www.arXiv.org/abs/1111.1987}{{\tt 1111.1987}}.

\bibitem{Gaberdiel:2011aa}
M.~R. Gaberdiel and P.~Suchanek, ``{Limits of Minimal Models and Continuous
  Orbifolds},'' {\em JHEP} {\bf 1203} (2012) 104,
\href{http://www.arXiv.org/abs/1112.1708}{{\tt 1112.1708}}.

\bibitem{Candu:2012jq}
C.~Candu and M.~R. Gaberdiel, ``{Supersymmetric holography on $AdS_3$},''
\href{http://www.arXiv.org/abs/1203.1939}{{\tt 1203.1939}}.

\bibitem{Candu:2012tr}
C.~Candu and M.~R. Gaberdiel, ``{Duality in N=2 Minimal Model Holography},''
  {\em JHEP} {\bf 1302} (2013) 070,
\href{http://www.arXiv.org/abs/1207.6646}{{\tt 1207.6646}}.

\bibitem{Candu:2012ne}
C.~Candu, M.~R. Gaberdiel, M.~Kelm, and C.~Vollenweider, ``{Even spin minimal
  model holography},'' {\em JHEP} {\bf 1301} (2013) 185,
\href{http://www.arXiv.org/abs/1211.3113}{{\tt 1211.3113}}.

\bibitem{Candu:2013uya}
C.~Candu and C.~Vollenweider, ``{The N=1 algebra $W_\infty[\mu]$ and its
  truncations},''
\href{http://www.arXiv.org/abs/1305.0013}{{\tt 1305.0013}}.

\bibitem{Gaberdiel:2012uj}
M.~R. Gaberdiel and R.~Gopakumar, ``{Minimal Model Holography},''
\href{http://www.arXiv.org/abs/1207.6697}{{\tt 1207.6697}}.

\bibitem{Brown:1986nw}
J.~D. Brown and M.~Henneaux, ``{Central Charges in the Canonical Realization of
  Asymptotic Symmetries: An Example from Three-Dimensional Gravity},'' {\em
  Commun.Math.Phys.} {\bf 104} (1986)
207--226.

\bibitem{Drinfeld:1981aa}
G.~Drinfeld and V.~Sokolov {\em Sov. Math. Dokl.} {\bf 23} (1981) 457.

\bibitem{Drinfeld:1985aa}
V.~S. G.~Drinfeld {\em J. Sov. Math.} {\bf 30} (1985) 1975.

\bibitem{Henneaux:2010xg}
M.~Henneaux and S.-J. Rey, ``{Nonlinear $W_{infinity}$ as Asymptotic Symmetry
  of Three-Dimensional Higher Spin Anti-de Sitter Gravity},'' {\em JHEP} {\bf
  1012} (2010) 007,
\href{http://www.arXiv.org/abs/1008.4579}{{\tt 1008.4579}}.

\bibitem{Henneaux:2012ny}
M.~Henneaux, G.~Lucena~Gomez, J.~Park, and S.-J. Rey, ``{Super- W(infinity)
  Asymptotic Symmetry of Higher-Spin $AdS_3$ Supergravity},'' {\em JHEP} {\bf
  1206} (2012) 037,
\href{http://www.arXiv.org/abs/1203.5152}{{\tt 1203.5152}}.

\bibitem{Hanaki:2012yf}
K.~Hanaki and C.~Peng, ``{Symmetries of Holographic Super-Minimal Models},''
\href{http://www.arXiv.org/abs/1203.5768}{{\tt 1203.5768}}.

\bibitem{Campoleoni:2010zq}
A.~Campoleoni, S.~Fredenhagen, S.~Pfenninger, and S.~Theisen, ``{Asymptotic
  symmetries of three-dimensional gravity coupled to higher-spin fields},''
  {\em JHEP} {\bf 1011} (2010) 007,
\href{http://www.arXiv.org/abs/1008.4744}{{\tt 1008.4744}}.

\bibitem{Gaberdiel:2011wb}
M.~R. Gaberdiel and T.~Hartman, ``{Symmetries of Holographic Minimal Models},''
  {\em JHEP} {\bf 1105} (2011) 031,
\href{http://www.arXiv.org/abs/1101.2910}{{\tt 1101.2910}}.

\bibitem{Campoleoni:2011hg}
A.~Campoleoni, S.~Fredenhagen, and S.~Pfenninger, ``{Asymptotic W-symmetries in
  three-dimensional higher-spin gauge theories},'' {\em JHEP} {\bf 1109} (2011)
  113,
\href{http://www.arXiv.org/abs/1107.0290}{{\tt 1107.0290}}.

\bibitem{Feher:1992yx}
L.~Feher, L.~O'Raifeartaigh, P.~Ruelle, I.~Tsutsui, and A.~Wipf, ``{On
  Hamiltonian reductions of the Wess-Zumino-Novikov-Witten theories},'' {\em
  Phys.Rept.} {\bf 222} (1992)
1--64.

\bibitem{Bouwknegt:1992wg}
P.~Bouwknegt and K.~Schoutens, ``{W symmetry in conformal field theory},'' {\em
  Phys.Rept.} {\bf 223} (1993) 183--276,
\href{http://www.arXiv.org/abs/hep-th/9210010}{{\tt hep-th/9210010}}.

\bibitem{DickeyLectures}
L.~Dickey, ``Lectures on classical {W}-algebras,'' {\em Acta Applicandae
  Mathematica} {\bf 47} (1997), no.~3, 243--321.

\bibitem{Ammon:2011nk}
M.~Ammon, M.~Gutperle, P.~Kraus, and E.~Perlmutter, ``{Spacetime Geometry in
  Higher Spin Gravity},'' {\em JHEP} {\bf 1110} (2011) 053,
\href{http://www.arXiv.org/abs/1106.4788}{{\tt 1106.4788}}.

\bibitem{Polyakov:1989dm}
A.~M. Polyakov, ``{Gauge Transformations and Diffeomorphisms},'' {\em
  Int.J.Mod.Phys.} {\bf A5} (1990)
833.

\bibitem{Bershadsky:1990bg}
M.~Bershadsky, ``{Conformal field theories via Hamiltonian reduction},'' {\em
  Commun.Math.Phys.} {\bf 139} (1991)
71--82.

\bibitem{Bilal:1991cf}
A.~Bilal, ``{W algebras from Chern-Simons theory},'' {\em Phys.Lett.} {\bf
  B267} (1991)
487--496.

\bibitem{Bilal:1992ub}
A.~Bilal, ``{All W(3) algebras from SL(3) Chern-Simons theory},'' {\em
  Phys.Lett.} {\bf B279} (1992)
308--318.

\bibitem{Banados:1992wn}
M.~Banados, C.~Teitelboim, and J.~Zanelli, ``{The Black hole in
  three-dimensional space-time},'' {\em Phys.Rev.Lett.} {\bf 69} (1992)
  1849--1851,
\href{http://www.arXiv.org/abs/hep-th/9204099}{{\tt hep-th/9204099}}.

\bibitem{Banados:1992gq}
M.~Banados, M.~Henneaux, C.~Teitelboim, and J.~Zanelli, ``{Geometry of the
  (2+1) black hole},'' {\em Phys.Rev.} {\bf D48} (1993) 1506--1525,
\href{http://www.arXiv.org/abs/gr-qc/9302012}{{\tt gr-qc/9302012}}.

\bibitem{Gutperle:2011kf}
M.~Gutperle and P.~Kraus, ``{Higher Spin Black Holes},'' {\em JHEP} {\bf 1105}
  (2011) 022,
\href{http://www.arXiv.org/abs/1103.4304}{{\tt 1103.4304}}.

\bibitem{Kraus:2012uf}
P.~Kraus and E.~Perlmutter, ``{Probing higher spin black holes},'' {\em JHEP}
  {\bf 1302} (2013) 096,
\href{http://www.arXiv.org/abs/1209.4937}{{\tt 1209.4937}}.

\bibitem{Perez:2012cf}
A.~Perez, D.~Tempo, and R.~Troncoso, ``{Higher spin gravity in 3D: black holes,
  global charges and thermodynamics},''
\href{http://www.arXiv.org/abs/1207.2844}{{\tt 1207.2844}}.

\bibitem{Campoleoni:2012hp}
A.~Campoleoni, S.~Fredenhagen, S.~Pfenninger, and S.~Theisen, ``{Towards
  metric-like higher-spin gauge theories in three dimensions},''
\href{http://www.arXiv.org/abs/1208.1851}{{\tt 1208.1851}}.

\bibitem{Perez:2013xi}
A.~Perez, D.~Tempo, and R.~Troncoso, ``{Higher spin black hole entropy in three
  dimensions},''
\href{http://www.arXiv.org/abs/1301.0847}{{\tt 1301.0847}}.

\bibitem{Kraus:2013esi}
P.~Kraus and T.~Ugajin, ``{An Entropy Formula for Higher Spin Black Holes via
  Conical Singularities},''
\href{http://www.arXiv.org/abs/1302.1583}{{\tt 1302.1583}}.

\bibitem{deBoer:2013gz}
J.~de~Boer and J.~I. Jottar, ``{Thermodynamics of Higher Spin Black Holes in
  AdS$_{3}$},''
\href{http://www.arXiv.org/abs/1302.0816}{{\tt 1302.0816}}.

\bibitem{Ferlaino:2013vga}
M.~Ferlaino, T.~Hollowood, and S.~P. Kumar, ``{Asymptotic symmetries and
  thermodynamics of higher spin black holes in AdS3},''
\href{http://www.arXiv.org/abs/1305.2011}{{\tt 1305.2011}}.

\bibitem{David:2012iu}
J.~R. David, M.~Ferlaino, and S.~P. Kumar, ``{Thermodynamics of higher spin
  black holes in 3D},'' {\em JHEP} {\bf 1211} (2012) 135,
\href{http://www.arXiv.org/abs/1210.0284}{{\tt 1210.0284}}.

\bibitem{Banerjee:2012aj}
S.~Banerjee, A.~Castro, S.~Hellerman, E.~Hijano, A.~Lepage-Jutier, {\em et
  al.}, ``{Smoothed Transitions in Higher Spin AdS Gravity},''
\href{http://www.arXiv.org/abs/1209.5396}{{\tt 1209.5396}}.

\bibitem{Chen:2012ba}
B.~Chen, J.~Long, and Y.-N. Wang, ``{Phase Structure of Higher Spin Black
  Hole},'' {\em JHEP} {\bf 1303} (2013) 017,
\href{http://www.arXiv.org/abs/1212.6593}{{\tt 1212.6593}}.

\bibitem{Ammon:2012wc}
M.~Ammon, M.~Gutperle, P.~Kraus, and E.~Perlmutter, ``{Black holes in three
  dimensional higher spin gravity: A review},''
\href{http://www.arXiv.org/abs/1208.5182}{{\tt 1208.5182}}.

\bibitem{Gaberdiel:2012yb}
M.~R. Gaberdiel, T.~Hartman, and K.~Jin, ``{Higher Spin Black Holes from
  CFT},'' {\em JHEP} {\bf 1204} (2012) 103,
\href{http://www.arXiv.org/abs/1203.0015}{{\tt 1203.0015}}.

\bibitem{Boussinesq:1872}
J.~Boussinesq, ``{Th\'eorie des ondes et des remous qui se propagent le long d'
  un canal rectangulaire horizontal, en communiquant au liquide contenu dans ce
  canal des vitesses sensiblement pareilles de la surface au fond},'' {\em
  Journal de Math\'ematiques Pures et Appliqu\'ees} {\bf 17(2)} (1872) 55--108.

\bibitem{Zacharov:1974}
V.~E. Zacharov, ``{On stochastization of one-dimensional chains of nonlinear
  oscilators},'' {\em Sov. Phys. JETP} {\bf 38(1)} (1974) 108--110.

\bibitem{McKean:1981aa}
H.~P. McKean, ``Boussinesq's equation on the circle,'' {\em Communications on
  Pure and Applied Mathematics} {\bf 34} (1981), no.~5, 599--691.

\bibitem{Deift:1982}
P.~Deift, C.~Tomei, and E.~Trubowitz, ``{ Inverse scattering and the Boussinesq
  equation},'' {\em Comm. on Pure and Appl. math.} {\bf 35(5)} (1982) 567--628.

\bibitem{Mathieu:1988pm}
P.~Mathieu, ``{Extended Classical Conformal Algebras and the Second Hamiltonian
  Structure of Lax Equations},'' {\em Phys.Lett.} {\bf B208} (1988)
101.

\bibitem{Mathieu:1991et}
P.~Mathieu and W.~Oevel, ``{The W(3)(2) conformal algebra and the Boussinesq
  hierarchy},'' {\em Mod.Phys.Lett.} {\bf A6} (1991)
2397--2410.

\bibitem{Bilal:1991dk}
A.~Bilal, ``{W algebras from Chern-Simons theory. 2},'' {\em Phys.Lett.}
(1991).

\bibitem{Zamolodchikov:1985aa}
A.~B. Zamolodchikov, ``{Infinite additional symmetries in two-dimensional
  conformal quantum field theory},'' {\em Theoret. Math. Phys.} {\bf 65} (1985)
  1205.

\bibitem{deBoer:1993iz}
J.~de~Boer and T.~Tjin, ``{The Relation between quantum W algebras and Lie
  algebras},'' {\em Commun.Math.Phys.} {\bf 160} (1994) 317--332,
\href{http://www.arXiv.org/abs/hep-th/9302006}{{\tt hep-th/9302006}}.

\bibitem{Castro:2012bc}
A.~Castro, E.~Hijano, and A.~Lepage-Jutier, ``{Unitarity Bounds in AdS$_3$
  Higher Spin Gravity},'' {\em JHEP} {\bf 1206} (2012) 001,
\href{http://www.arXiv.org/abs/1202.4467}{{\tt 1202.4467}}.

\bibitem{Batlle:1992uu}
C.~Batlle, ``{Lecture notes on KdV hierarchies and pseudodifferential
  operators},'' {\em Unpublished} (1992)
  \href{http://www.arXiv.org/abs/http://www-ma4.upc.edu/\~carles/listpub.html}{{\tt
  http://www-ma4.upc.edu/\~carles/listpub.html}}.

\bibitem{Dickey:1997aa}
L.~Dickey, ``{Lectures on Classical W-Algebras},'' {\em Acta Applicandae
  Mathematica} {\bf 47} (1997), no.~3, 243--321.

\bibitem{Gelfand:1987aa}
I.~GelÕfand and L.~Dickey, ``{},'' {\em Annals of the New York Academy of
  Science} {\bf 491} (1987) 131.

\bibitem{Barnich:2007bf}
G.~Barnich and G.~Comp\`ere, ``{Surface charge algebra in gauge theories and
  thermodynamic integrability},'' {\em J.Math.Phys.} {\bf 49} (2008) 042901,
\href{http://www.arXiv.org/abs/0708.2378}{{\tt 0708.2378}}.

\bibitem{Hartman:2008dq}
T.~Hartman and A.~Strominger, ``{Central Charge for AdS(2) Quantum Gravity},''
  {\em JHEP} {\bf 0904} (2009) 026,
\href{http://www.arXiv.org/abs/0803.3621}{{\tt 0803.3621}}.

\bibitem{Malik:1996rb}
R.~Malik, ``{Commuting conserved quantities in nonlinear realizations of
  W(3)},'' {\em Phys.Lett. B}
(1996).

\bibitem{DeGroot:1991ca}
M.~F. De~Groot, T.~J. Hollowood, and J.~L. Miramontes, ``{Generalized
  Drinfeld-Sokolov hierarchies},'' {\em Commun.Math.Phys.} {\bf 145} (1992)
57--84.

\bibitem{Burroughs:1991bd}
N.~J. Burroughs, M.~F. de~Groot, T.~J. Hollowood, and J.~L. Miramontes,
  ``{Generalized Drinfeld-Sokolov hierarchies 2: The Hamiltonian structures},''
  {\em Commun.Math.Phys.} {\bf 153} (1993) 187--215,
\href{http://www.arXiv.org/abs/hep-th/9109014}{{\tt hep-th/9109014}}.

\end{thebibliography}

\providecommand{\href}[2]{#2}\begingroup\raggedright\endgroup

\end{document}